**Physics Oriented Mathematical Perspective for Creating Trochoids and Ellipses through the Combination Rolling and Sliding Motions of a Circle Along Another One in Forward and Backward Sliding Modes**

H. arbab[1] , Physics department, university of Kashan, Kashan, Iran

Arzhang. arbab[2], student of mathematics, university of Kashan, Kashan, Iran

**Abstract**

*Usual mathematical method for creating trochoids is based on a solid rule that requires a pure rolling motion of a circle along another one. In this vision a trochoid defined as a traced path by an attached point (a non- conceptive issue) to a pure rolling circle! Except for using the restriction of pure rolling motion for a circle beside using the attached points to it, authors of this article have not found other conceptive solutions for this issue in the references of mathematics and physics. This article provides a novel conceptive solution for creating trochoids and ellipses based on combination of rolling and sliding motions of a circle along another one! Therefore, we have not to define a trochoid as a path that is swept by an attached point to a pure rolling circle along another circle! Instead, a trochoid can be defined as a path is swept by a point on the circumference of a circle that is doing rolling and sliding motions uniformly along another one! Also, this article presents two different methods to implement a mathematical simulation for a moving circle does uniform simultaneous rolling and sliding motions along another one! With the help of this innovative solution, it is possible to define ellipses and trochoids as closed plane curves that can be generated through the combination of rolling and sliding motions [ellipse is created through the combination of two co-polarized rotational motions with different commensurable angular frequencies (in two different modes!)]. This article presents a novel idea titled by Virtual Rotating Circles Technique (VRCT) that can be implemented by Mathematical Simulator Machine.*

**1A. Introduction** As we know a path that is swept by an attached point to a pure rolling circle along another circle is a centered trochoid. *Search in references of physics and mathematics show development on the subject of trochoid family is based on a non-conceptive rigid rule [2-9] and issue of attached points to pure rolling motion of a circle along another one. On the other hand, there is not found another conceptive description to show that process of creating ellipses in the framework of creating hypotrochoids family. In addition, creating ellipse through the superposition of two perpendicular simple harmonic oscillations (with same frequencies) is a usual process in physics [1] but in this article, we are going to show that not only an ellipse can be created through the combination of co-polarized rotational motions with different commensurable angular frequencies but also it can be interpreted as a uniform combination of rolling and sliding motions of a circle inside another one!* Investigation in mathemetics and physics sources show that there is no established mental vision prove that centered trochoids and ellipses can be created through the combination of rolling and sliding motions of a circle along another one in two modes of sliding motions! Authors have not successed to find a mathematical or physical reference to show that how the parametric equations of centered trochoids must be derived on the bases of simultaneous rolling and sliding motions of a circle along another one. *In this article we want to present a conceptive method for creating trochoids and ellipses through a common framework based on the combination of rolling and sliding motions of a circle $C_b$ along another circle $C_{a\pm b}$.Only* epicycloid and hypocycloid are especial types of geometrical objects traced by a point p on the circumference of a pure rolling circle along another one [3]. In the presented references, other geometrical objects in the trochoids family are considered as the traced paths of certain points attached to pure rolling circles. Theoretical analysis about creating trochoids through pure rolling motion of a circle beside the subject of attached points is not a complicated and challenging mathematical issue. However, implementing a comprehensive and observable method for creating trochoids and ellipses through the combination of rolling and sliding motions of a circle along another circle seems to be a very complex and challenging issue. *The first serious challenging issue for creating trochoids through the combination of rolling and sliding motions is due to its technical process (implementing a practically method for making uniform-controlled combination of rolling and sliding motions). The second challenging issue is come back to the fact that an efficient mathematical method has not been discovered to deduce the parametric equations of the path that is swept by a certain point on the circumference of a rolling and sliding circle along another one*. Deduction parametric equations of epicycloid or hypocycloid (as the paths that are swept by a definite point on the circumference of the pure rolling circle) is a simple mathematical issue rather than the same issue when a circle is rolling and sliding simultaneously along another circle. Implement a pure rolling situation can be met by using different kinds of gears as is done by a spirograph [10, 11]. In the other words, a method has not been devised yet to describe other geometrical objects in the trochoids family as the objects that are swept by a specific point on the circumference of a rolling and sliding circle. In this article, we are also going to develop conceptive applicable methods to implement a uniform controllable combination of rolling and sliding motions for a moving circle along another one. Such a method can be performed simply in laboratory with the help of mathematical simulator machine. Through the analysis about combination of rolling and sliding motions of a circle (along another circle or a straight line) professional staffs would be able to implement our conceptive and observable methods for combination of rolling and sliding motions. Therefore, on the bases of the method which is presented here, there is no need to consider an extended spectrum of trochoids family as the paths where are traced by attached points to pure rolling circles! With the help of mathematical simulator machine, it would be possible to create all of the trochoid's family and co-centered ellipses (including line segments and circles) in two different modes of sliding motion. *On the bases of new vision that is presented here, all the trochoids family and ellipses can be considered as the paths of a definite point on the circumference of circles that are rolling and sliding uniformly along another circles! Then, deduction parametric equations of a path that is swept by a point on the circumference of a rolling and sliding circle (along another one) would not be a strange issue. Without using the method that is presented in this article, it seems description the subject is impossible (or at least is a very challenging mathematical issue).* Understanding the contents of this paper can help the readers to change their usual mathematical vision about creating centered trochoids (and ellipses) that is constructed on the bases of a non-conceptive rigid rule. With the help of our new perspective, the constructive role of physical concepts (such as frequency, polarization, phase, etc) come in the mathematics of trochoids and ellipses. Therefore, at this moment this article establishes a strong link between mathematics and realities of the physical world. In the other words this article presents is a new marriage of physics and mathematics.

[1] arbabpen@kashanu.ac.ir, corresponding author, +98 09136995136

[2] aarbab2022@yahoo.com

**1B. Mathematical Simulation of Rolling a Circle Along Another One**

It is clear that implement the pure rolling motion of a circle $C_b$ (with radius b) along another circle $C_{a\pm b}$ (with radius $a \pm b$ )without sliding is a challenging experimental process in laboratory unless we use gears instead usual circles in a device such as spirograph. (Parameter a stands for distance between the center of circles $C_b$ and $C_{a\pm b}$ as shown by Figs 1a-1d. In this article, we will be dealing with a more challenging problem of using combination of rolling and sliding motions for a circle $C_b$ along $C_{a\pm b}$ . This issue would be more complicated if we have to change radius of circles continuously. Since, even in the case of simple pure rolling issue using different gears instead of $C_b$ and $C_{a\pm b}$ beside changing radiuses continuously would not be possible experimentally. Implement simultaneous rolling and sliding motions of a circle (that we will be facing in this article beside changing radiuses continuously) is even would be more complicated than the establish pure rolling issue. Therefore, in order to implement the process of rolling (or rolling and sliding motions) for a circle $C_b$ along another circle $C_{a\pm b}$, we need an intelligent mathematical perspective to create a practical method to do this complicated issue. Thus, the first important issue that we are facing with, is finding a simple method for implement the pure rolling motion of an arbitrary circle along another selective one. In order to overcome the problem of combination rolling and sliding motions we will present a novel idea to consider a rotating tablet (about its symmetry axis) as a continuous spectrum of the co-centered virtual rotating circles. Therefore, in the first step we must present an important theorem for pure rolling motion of a circle along another one under the title: "Simulating Rolling Circles Theorem" (SRCT) as follows:

**Theorem1**-Assume that a circle $C_b$ has pure rolling motion along a fixed circle $C_{a\mp b}$ as shown by Fig1a and Fig1c respectively. Then, the traced path by a point p (on the circumference of $C_b$) in the fixed coordinate system XOY (shown by Fig1a and Fig1c) is same as the traced path by point p in a coordinate system XOY (in Fig1b and Fig1d) that is rotating with angular frequency $\Omega = b\omega/(a \mp b)$ respectively. [Note that centers of circles $C_b$ and $C_{a\mp b}$ (in Fig1b and Fig1d) are fixed on $X^*$ while the circles are rotating as shown by the figures under consideration about their centers. Also, coordinate systems XOY (in Fig1b and Fig1d) are rotating in-phase along with the circles $C_{a\mp b}$]

Proof: In the first step we consider Fig1a for a pure rolling motion of circle $C_b$ along the circle $C_{a-b}$. Consider a point p on the circumference of rolling circle $C_b$. It is assumed that at the starting time $t = 0$ the position of point p is set on the common axis $X\,X^*$ . To prove the theorem1 we must show that equation of the path is swept by point p on the fixed coordinate system XOY (in Fig1a) is same as the equation of the traced path by point p in the rotating coordinate system XOY (in Fig1b).

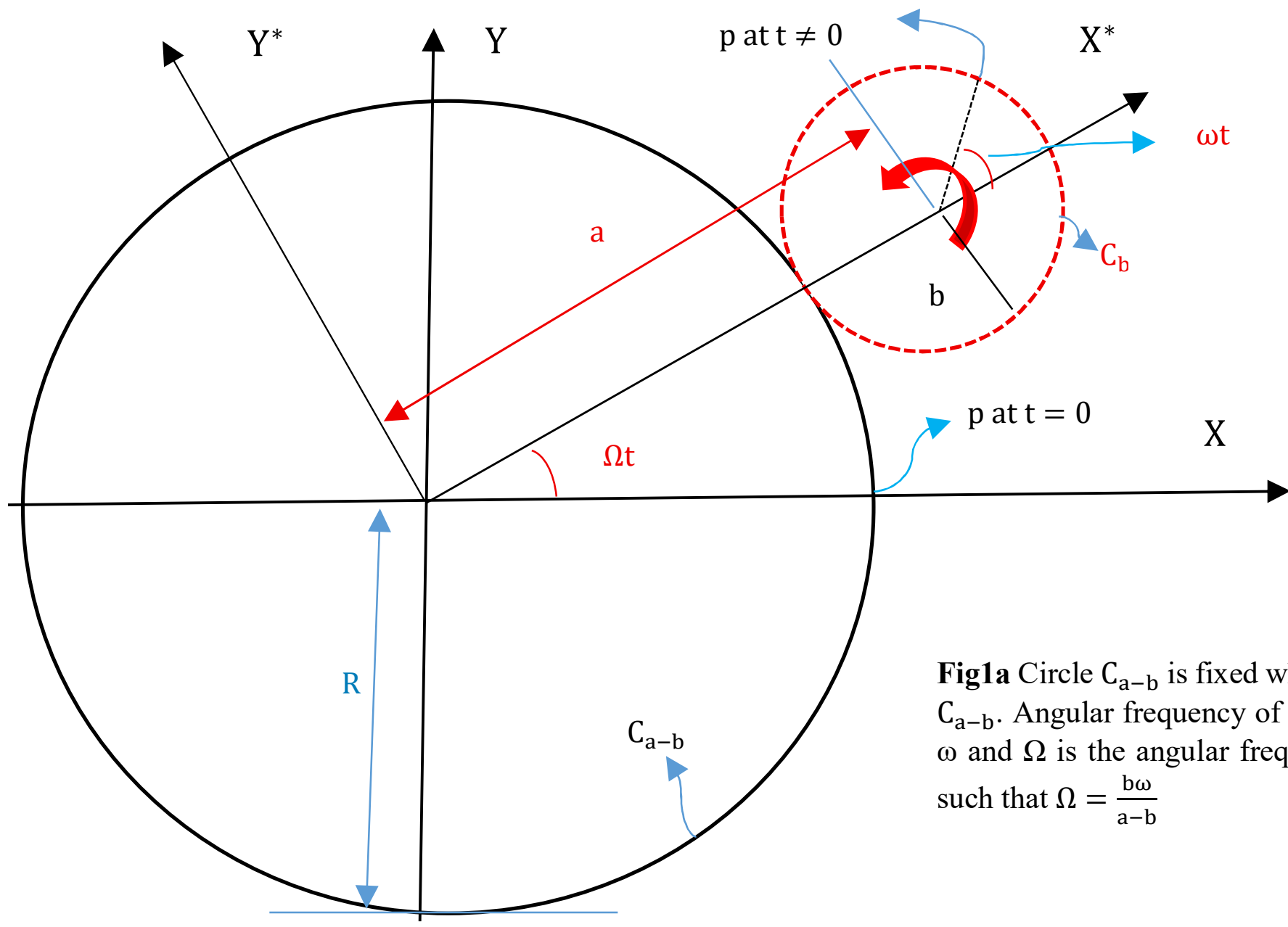

**Fig1a** Circle $C_{a-b}$ is fixed while Circle $C_b$ rolls outside the circle $C_{a-b}$. Angular frequency of rolling circle $C_b$ around its center is ω and Ω is the angular frequency of $C_b$ around the circle $C_{a-b}$ such that $\Omega = \frac{b\omega}{a-b}$

Consider rolling circle $C_b$ that is contacted with a fixed position circle $C_{a-b}$ in Fig1a. The angular frequency of $C_b$ (due to its rolling motion along $C_{a-b}$) around its center to be ω. Pure rolling motion of $C_b$ requires that the angular frequency of rotating system $X^*Y^*$ to be $\Omega = \frac{b\omega}{a-b}$. Now consider a different situation that is shown by Fig1b. In this figure center of rotating circles $C_b$ and $C_{a-b}$ are fixed on $X^*$while they are in contact and rotating in opposite directions. Also, in this situation the position of point p is considered to be set on the common axis X $X^*$ at the starting time $t = 0$. Circle $C_b$ rotates with angular frequency ω around its center while the circle $C_{a-b}$ (along with coordinate system XOY) rotates with angular frequency $\Omega = \frac{b\omega}{a-b}$ in the opposite direction as is shown by Fig1b. With the help of this mathematical simulator, along with the constraint $\Omega = b\omega/(a-b)$ it can be shown that the traced path of point p in rotating plane of XOY in Fig1b would be an epicycloid same as the traced path by point p in coordinate system XY (in Fig1a).

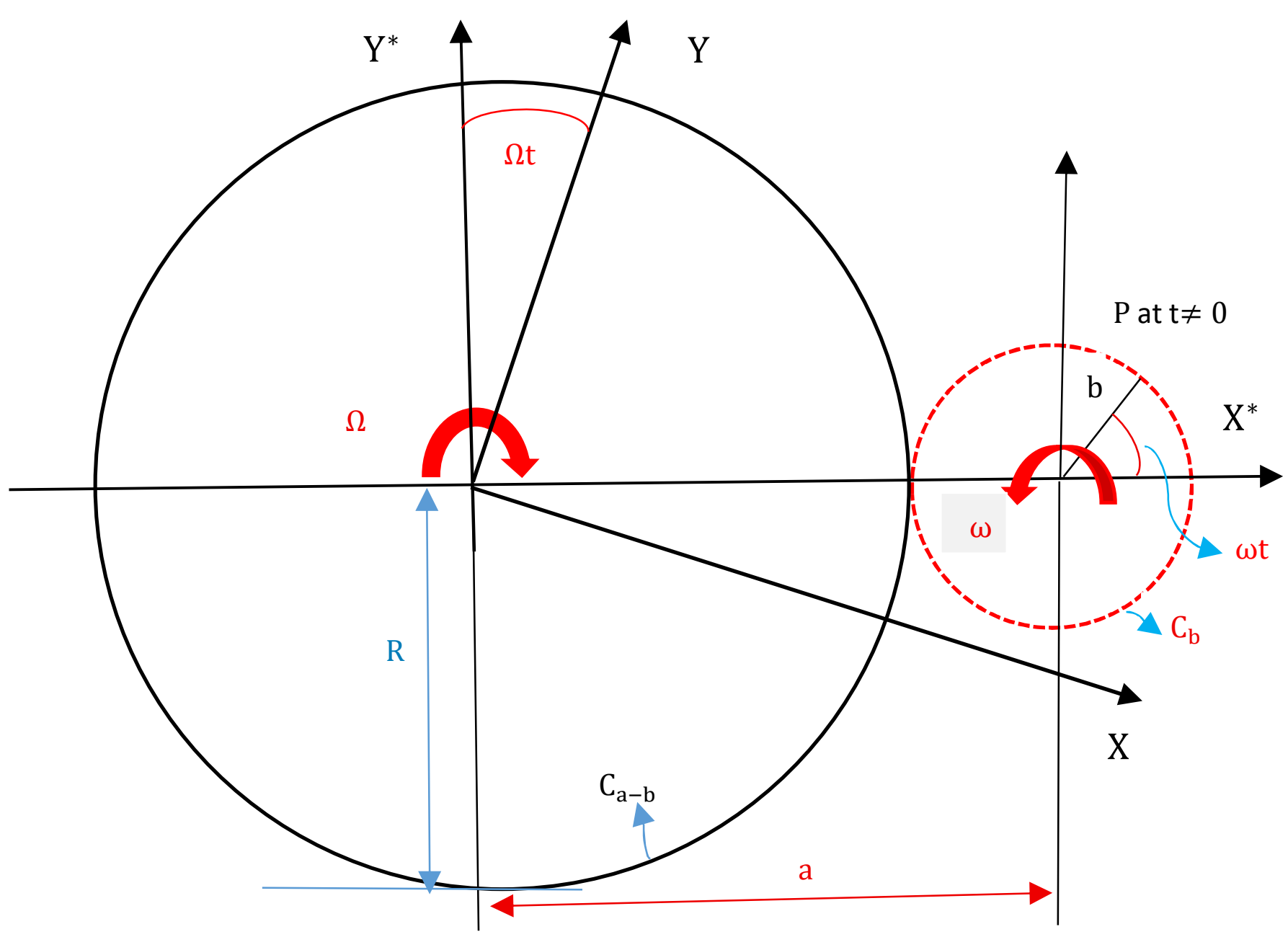

**Fig1b**. circles $C_b$ and $C_{a-b}$ are rotating around their symmetry axis with angular frequencies ω and Ω (in opposite directions) respectively. This picture shows a simple mathematical simulation of rolling circle $C_b$ around the circle $C_{a-b}$ in Fig1*a* such that $\Omega = \frac{b\omega}{a-b}$

From Fig1a it is clear that

$X^* = a + b\cos\omega t$ (1)

$Y^* = b\sin\omega t$ (2)

Coordinate transformation requires that

$X = X^*\cos\Omega t - Y^*\sin\Omega t$ (3)

$Y = X^*\sin\Omega t + Y^*\cos\Omega t$ (4)

Therefore,

$X = a\cos\Omega t + b\cos(\omega + \Omega)t$ (5)

$Y = a\sin\Omega t + b\sin(\omega + \Omega)t$ (6)

These equations are parametric equations of the traced path by point p in the fixed coordinate system XOY in Fig1a. On the other hand, from Fig1b equations (1) and (2) are also satisfied. Then, with the help of transformations (3) and (4) we obtain the same results that are given by (5), (6). Therefore, the traced path of point p in Fig1a is same as the path traced by p in the rotating coordinate XOY is shown by Fig1b. Same argument can be extended for a situation that a circle $C_b$ has a pure rolling motion inside a circle $C_{a+b}$ as shown by Fig1c. (The only difference of this situation respect to the previous one is that the rotation of circles $C_b$ and $C_{a+b}$ must be in the same direction)

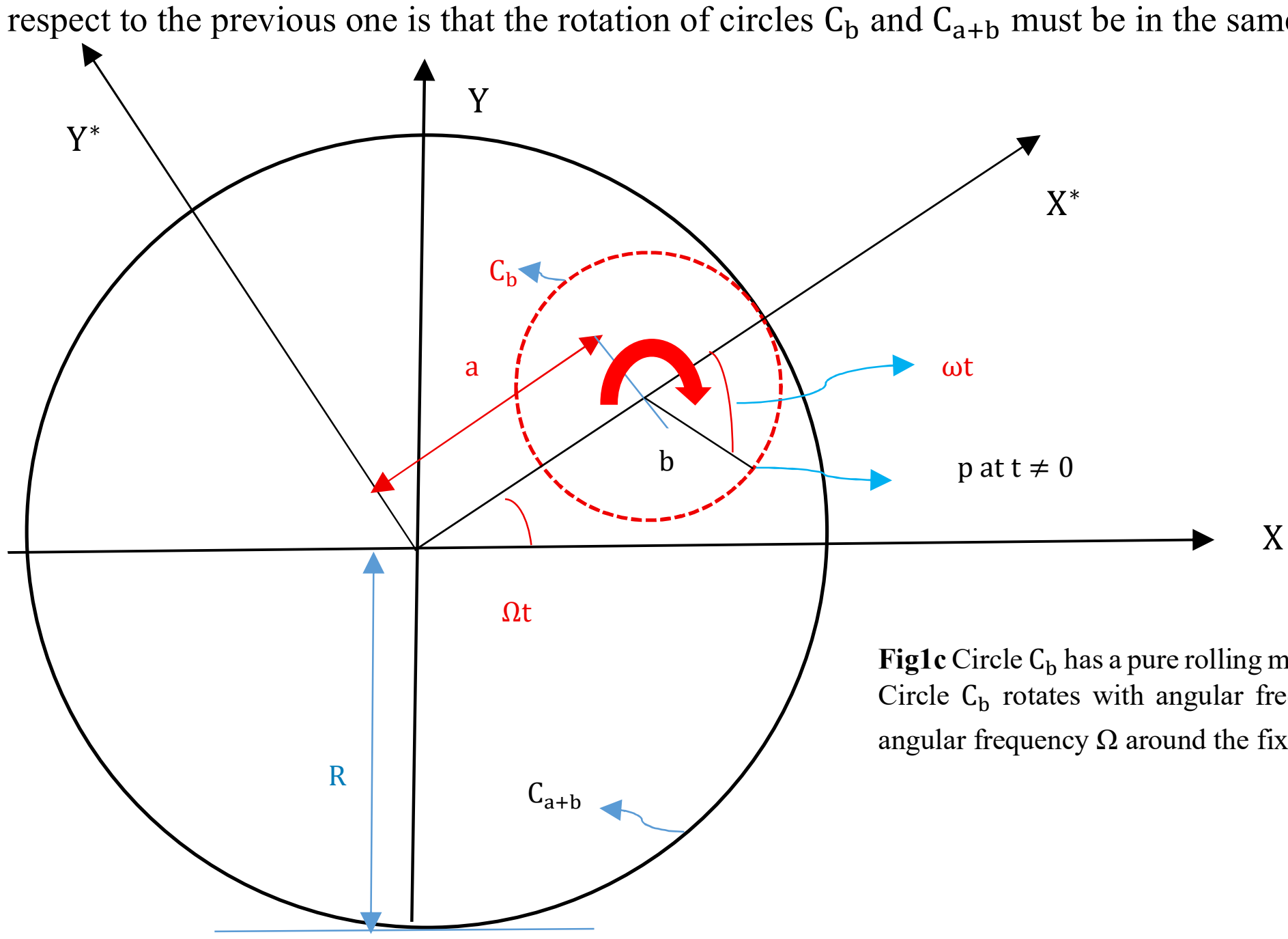

**Fig1c** Circle $C_b$ has a pure rolling motion inside the fixed circle $C_{a+b}$. Circle $C_b$ rotates with angular frequency ω around its center and angular frequency Ω around the fixed circle $C_{a+b}$ such that $\Omega = \frac{b\omega}{a+b}$

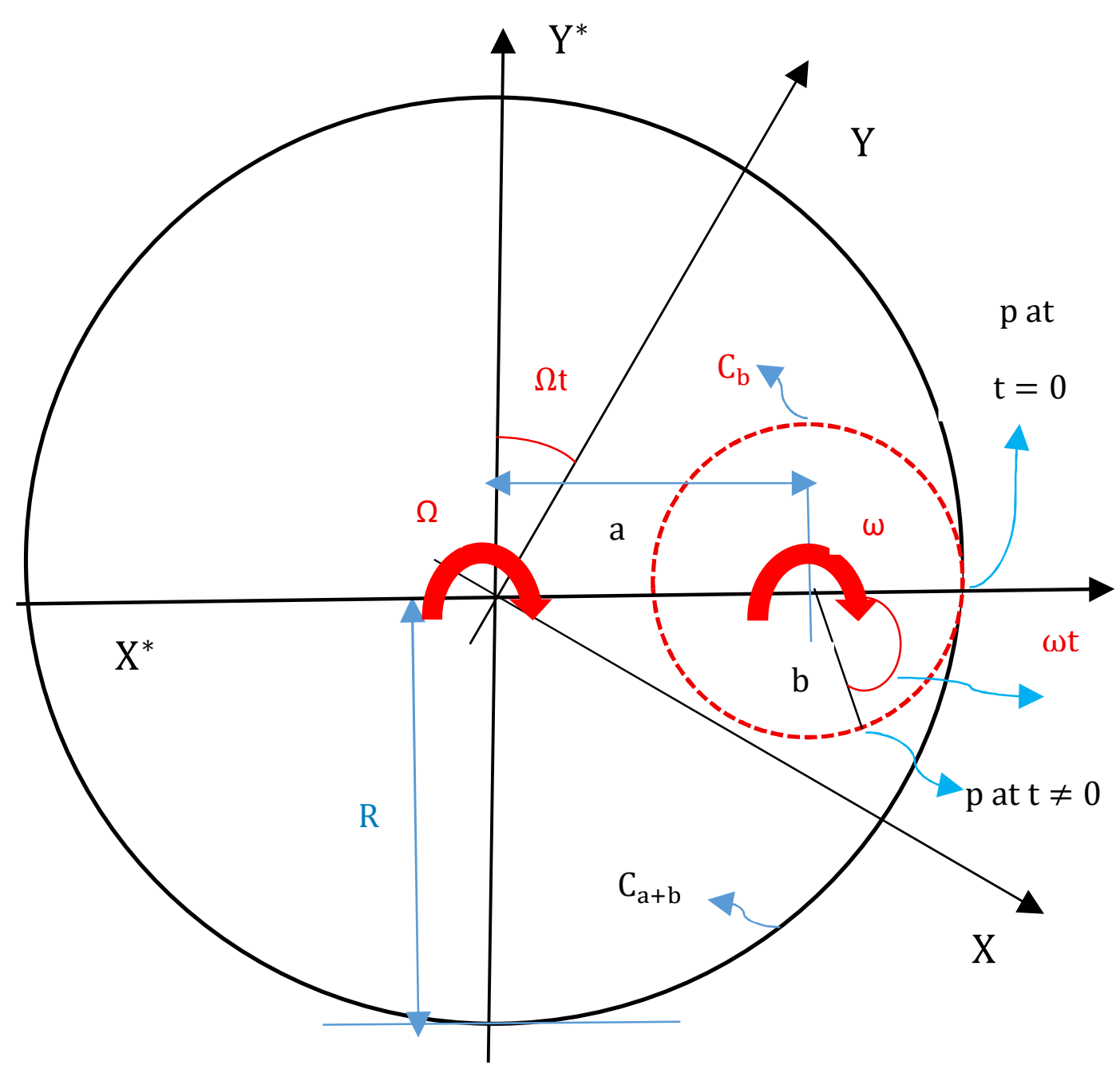

**Fig1d**. A simple mathematical simulator for Fig1c. Both circles $C_b$ and $C_{a+b}$ are rotating with angular frequencies ω and Ω respectively. Pure rolling motion of $C_b$ inside the circle $C_{a+b}$ in this mathematical simulator requires that $\Omega = \frac{b\omega}{a+b}$

From Fig1c we have

$$X^* = a + b\cos\omega t \tag{7}$$
$$Y^* = -b\sin\omega t \tag{8}$$
By using equations (3), (4), (7) and (8) we have
$$X = a\cos\Omega t + b\cos(\omega - \Omega)t \tag{9}$$
$$Y = a\sin\Omega t\ - b\sin(\omega - \Omega)t \tag{10}$$
On the other hand, from Fig1d it is clear that
$$X^* = a + b\cos\omega t \tag{11}$$
$$Y^* = -b\sin\omega t \tag{12}$$
Then, by using transformations (3) and (4) along with (11) and (12) we obtain again the same parametric equations (9) and (10) for traced path of p in rotating coordinate system XOY that is shown by Fig1d.This means that configuration that is shown by Fig1d simulates the process shown by Fig1c.
Now we assume that $\omega = \Omega$ , then, simulating pure rolling motion of a circle along another one with the help of Fig1b requires $a = 2b$ as shown by Fig1B. Also, assume that $a = b$ in Fig1d, then, simulating pure rolling motion requires that $\omega = 2\Omega$ as shown by Fig1D. In both last cases, when each of the requirements under consideration is violated then, the simulated process would be a combination of rolling and sliding motions. Thus, in such situations we introduce $\Delta v = b\omega - (a \mp b)\Omega$ as difference in linear speeds of two points on circumferences of $C_b$ and of $C_{a\mp b}$ respectively. Changing any of the parameters $a, b, \omega$ and $\Omega$ such that $\Delta v \neq 0$ leads us to violation the pure rolling requirement. Therefore, we must define two following modes of sliding motion:
$\Delta v < 0$ as **Forward sliding mode** and $\Delta v > 0$ as **Backward sliding mode.**

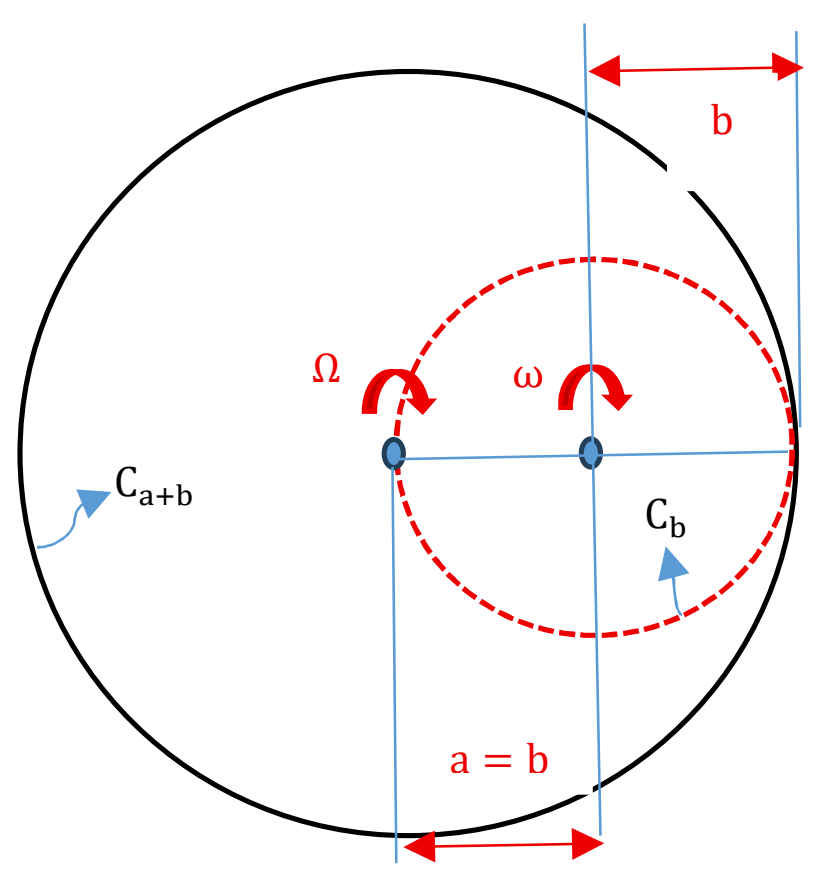

**Fig1D.** An especial case for Fig1d. Mathematical simulation for pure rolling motion of circle $C_b$ inside $C_{a+b}$ . $(\omega = 2\Omega\ , a = b)$

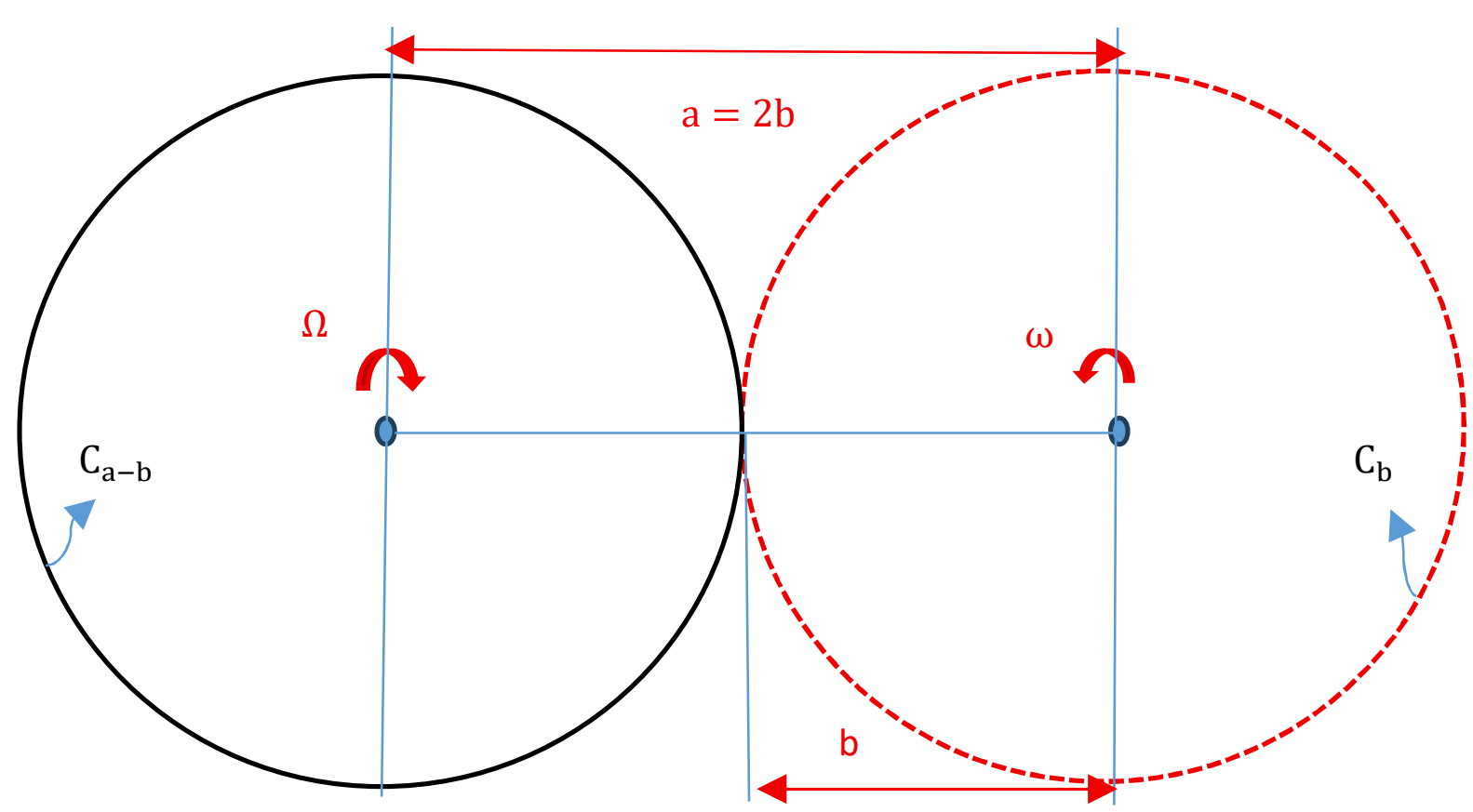

**Fig1B.** An especial version of Fig1b. Mathematical simulation for pure rolling motion of circle $C_b$ outside $C_{a-b}$ . $(\omega = \Omega\ , a = 2b)$

### 1C. Mathematical Simulator (MS)

Mathematical simulators are shown by Fig1b and Fig1d clear the outstanding role of physical concepts such as amplitude, frequency, phase, polarization, sliding and rolling motion in the field of geometry! Fig1e describes a sketch of an advanced innovative mathematical simulator machine that works on the based of combination rolling and sliding motions through VRCT. Therefore, instead of using a real physical rolling circle along another stationary circle it is possible to consider a continuous spectrum of virtual rotating circles on the plane of two rotating parallel tablets as shown by Fig1e! In order to apply a rolling (or uniform combination of rolling and sliding motions) for a virtual circle along another one, the distance between symmetry axes of the tablets, the distance between pen from symmetry axis of tablet 2, angular frequencies of the tablets and direction of rotation (polarization issue) of tablet 2 must be adjustable.

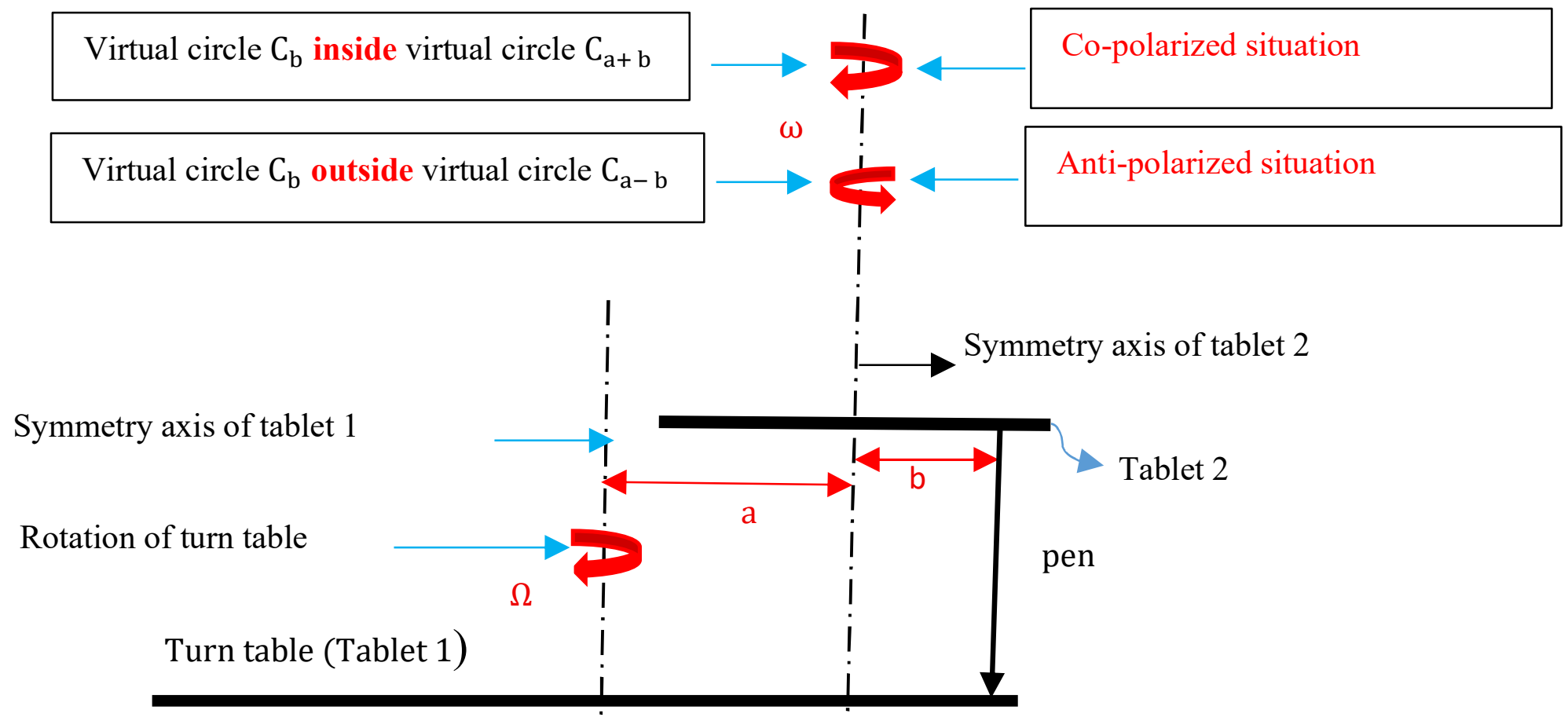

**Fig1e.** Two parallel rotating tablets are shown by bold black color. Tablets are perpendicular to the paper sheet. Pen is holding by tablet 2. Distance of pen with symmetry axis of the tablet 2 is adjustable. Also, distance between symmetry axis of the tablets must be adjustable. Direction of rotation of the tablet 2 determines the state of polarization. Rotation of tablets in same direction specifies co-polarized situation.

In order to establish a mathematical simulator (to create epitrochoids as shown by Fig1b), the tablets must be rotate in opposite directions while creation hypotrochoids or ellipses (as shown by Fig1d) both tablets must be rotate in same direction (polarization issue). Through the VRCT we are using two definite continuous spectrums of virtual circles $C_b$ on tablet 2 and $C_{a\pm b}$ on the tablet 1. Thus, by using a configuration such as shown by Fig1e we would be able to simulate a usual real rolling motion (or a uniform combination of rolling and sliding motions) of a virtual circle $C_b$ along another virtual circle $C_{a\pm b}$. Picture 1 shows MS along with an epicycloid and a hypocycloid that are plotted through VRCT.

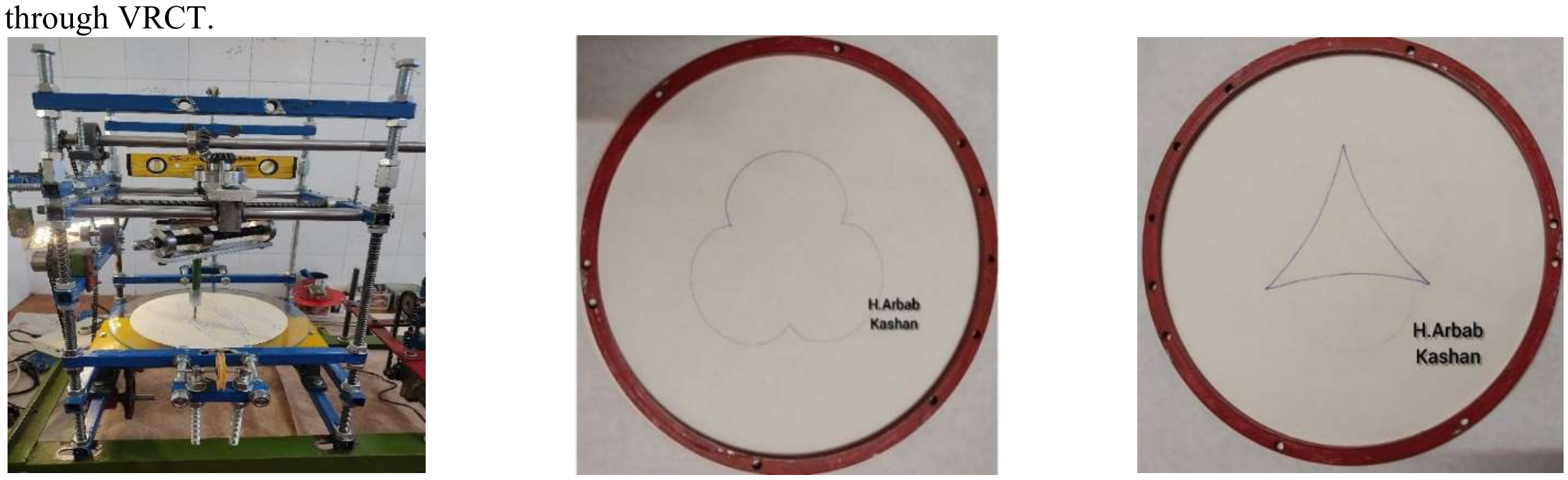

**Picture 1 Left side:** An advanced MS machine. Parallel tablets of the device can be rotating in same or opposite directions. The distance between symmetrical axis of the tablets and the position of pen on the tablet 2 are adjustable. **Right side:** An epicycloid and a hypocycloid plotted by MS machine.

As a result of theorem1, pure rolling motion of a circle $C_b$ inside/ outside the circle $C_{a\pm b}$ through VRCT, requires that the parameters under consideration to be adjusted such that

$$(a+b)\Omega = b\omega \quad (13)$$

$$(a-b)\Omega = b\omega \quad (14)$$

Respectively. Equations (13) and (14) guarantees the pure rolling motion through the (VRCT).

### 2A. Analytical Details for Creating an Epicycloid Through VRCT

According to definition, an epicycloid is a plane curve that is traced by a point p on the circumference of a pure rolling circle around a fixed stationary circle. Thus, in the first step in order to adjusting the device we must specify the polarization state, radius of virtual circles $C_b$ and $C_{a-b}$ on the rotating tablets along with the geometrical features of epicycloid. Fig2a help us to visualize and analyze the details of pure rolling process through VRCT in such a situation.

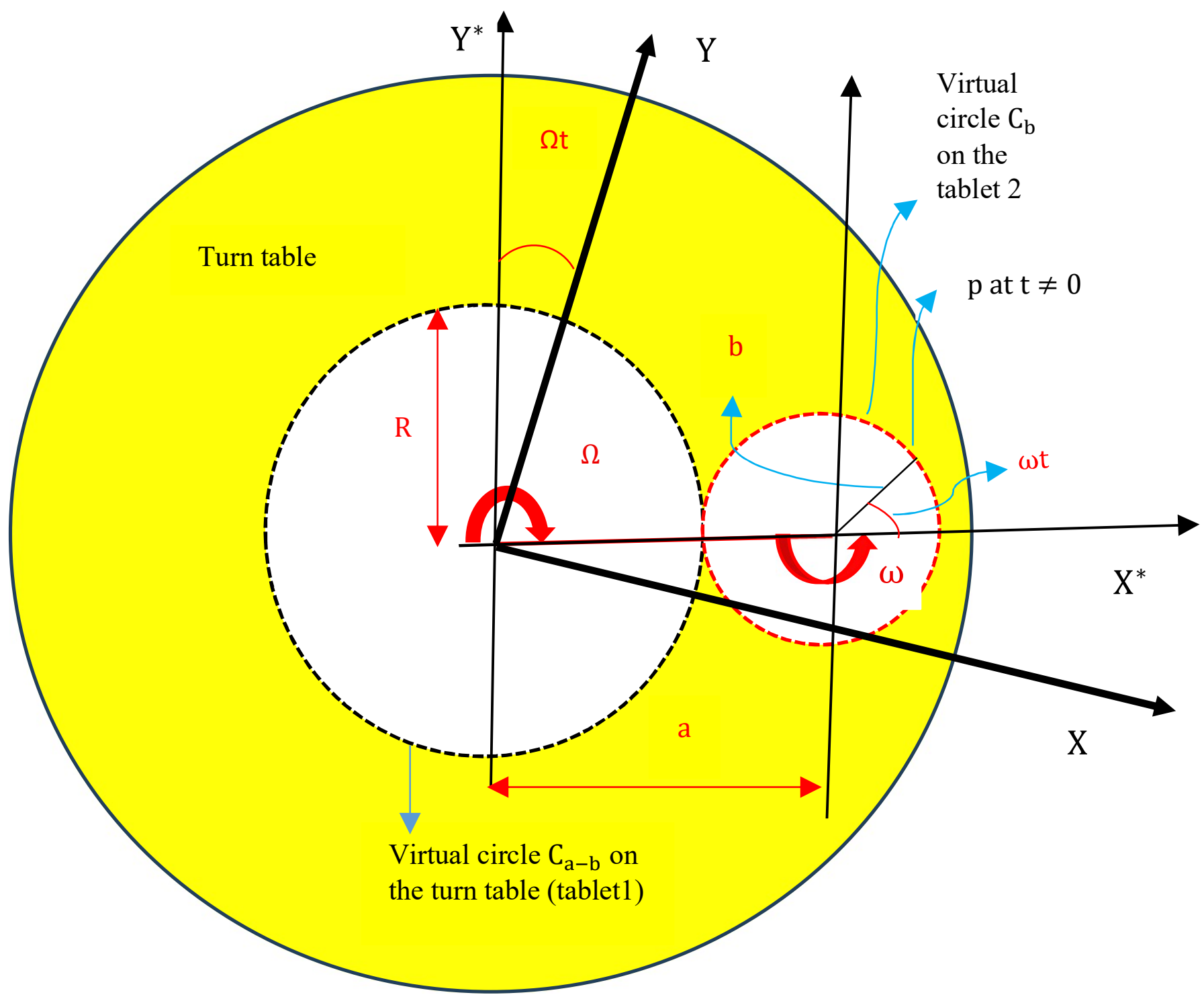

**Fig2a** Two virtual circles $C_b$ and $C_{a-b}$ (dashed lines) on the rotating tablets. XOY is rotating coordinate system that is fixed on the turn table. X*O*Y* is stationary coordinate system.

In order to use VRCT in the simplest case as shown by Fig2a, we assume that the tablets are rotating with equal angular frequencies. If $b = \frac{a}{2}$ then it follows that the virtual circles $C_b$ and $C_{a-b}$ will have same radiuses. Therefore, we expect the traced path of point p in the plane of rotating coordinate XOY to be a cardioid. By using equations (5), (6) and (14) we have

$$X = a\cos(\frac{b\omega}{a-b}t) + b\cos(\frac{a\omega}{a-b}t) = a\cos(\omega t) + \frac{a}{2}\cos(2\omega t) \quad (15)$$

$$Y = a\sin(\frac{b\omega}{a-b}t) + \frac{a}{2}\sin(\frac{a\omega}{a-b}t = a\sin(\omega t) + \frac{a}{2}\sin(2\omega t) \quad (16)$$

We take $a = 100$ then, for any integer or rational number ω the traced path of point p would be a cardioid. In general, suppose that the tablet 1 rotates with an angular frequency Ω and the tablet 2 rotates with angular frequency ω such that equation (14) is satisfied. Now, if the ratio $\frac{a-b}{b} = n_e$ to be an integer, then an epicycloid [contains $n_e$ repeated sections in one complete rotation of tablet1(turn table)] would be created. In the other words

$$b = \frac{a}{n_e + 1} \qquad n_e = 1,2,3,\ldots \quad (17)$$

Therefore, in order to form an epicycloid with $n_e$ repeated parts in one complete rotation of tablet 1 with a selected value for a (the distance between the centers of virtual circles $C_b$ and $C_{a-b}$ ) we must choose b on the bases of equation (17). Equation (17) along with the Fig2a require that

$$R = a - b = \frac{n_e a}{n_e + 1} \quad (18)$$

**Example 1.** Suppose that an epicycloid is created on the bases of VRCT and it has five repeated parts in one complete rotation of tablet1. If $a = 100$ , $\omega = 1$ , compute the value of b and Ω . In the first step by using (17) the value of b is determied. Thus, $b = \frac{100}{5+1} = \frac{50}{3}$ . To compute Ω , equation (14) beside the last result for b require $\Omega = \frac{1}{5}$. By using these results for b and Ω along with equations (5) , (6) we have

$$X = 100\cos(\frac{1}{5}t) + \frac{50}{3}\cos(\frac{6}{5}t) \quad (19)$$

$$Y = 100\sin(\frac{1}{5}t) + \frac{50}{3}\sin(\frac{6}{5}t) \tag{20}$$

Through replacment $+$ sign by $-$ sign in (20) the epicicloid under consideration convert to a hypocycloid (Fig2b).

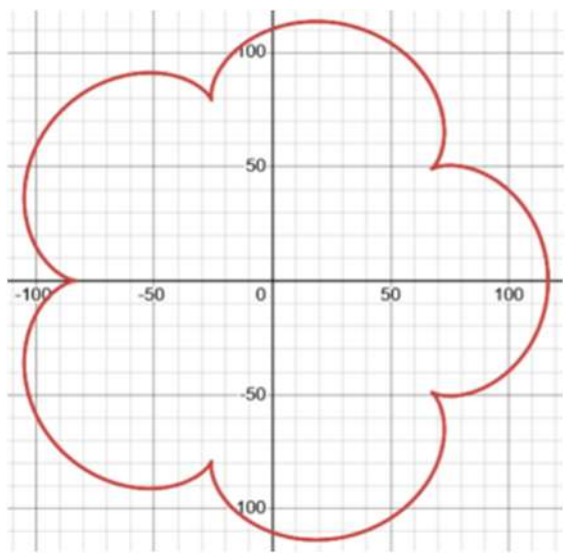

**Fig2b. Left side:** An epicycloid with five repeated parts in one complete rotation of tablet1 **Right side**: Hypocycloid that is created through the replacment $+$ sign by $-$ in (20). Both geometrical objects are plotted on the bases VRCT with the help of computer. This subject reflects the important role of polarization issue in crating centered trochoids

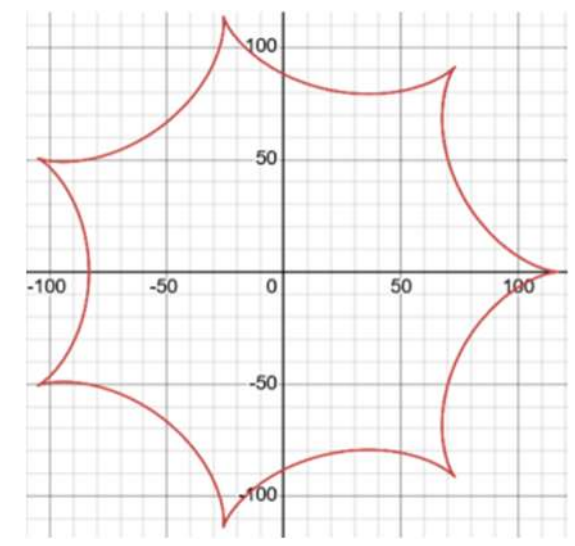

**2B. Uniform Combination of Rolling and Sliding Motions through Changing Parameter (UCRSMCP)**

*So far, we have focused on pure rolling motion of virtual circle $C_b$ along another virtual circle $C_{a-b}$ through the VRCT. At this point we are trying to introduce a fantastic method for doing uniform combination of rolling and sliding motion by virtual circles under consideration with the help of VRCT.* For definite values of $a$, $b$, $\omega$ and $\Omega$ leads us to an epicycloid, now it is possible to select different values for b such that equation (14) does not hold. In such a situation, linear speed of the virtual circles $C_b$ *and* $C_{a-b}$ *on* the contact point of the virtual circles *would be different. Therefore, we would have uniform combination of rolling and sliding motions simultaneously.* Consider the fallowing numerical example:

**Example 2.** Suppose that all the parameters (except b) are same as those that were given in example1. If we assume that $b = 30 > \frac{50}{3}$, $a = 100$, $\omega = 1$, $\Omega = \frac{1}{5}$ then, obviously with this new parameter for b, equation (14) does not hold. By replacement these parameters into equations (19 and (20)) we obtain following parametric equations:

$$X = 100\cos(\frac{1}{5}\omega t) + 30\cos(\frac{6}{5}\omega t) \tag{21}$$

$$Y = 100\sin(\frac{1}{5}\omega t) + 30\sin(\frac{6}{5}\omega t) \tag{22}$$

Also, if we take $b = 10 < \frac{50}{3}$, $a = 100$, $\omega = 1, \Omega = \frac{1}{5}$ then equations (19) and (20) would be replaced by

$$X = 100\cos(\frac{1}{5}\omega t) + 10\cos(\frac{6}{5}\omega t) \tag{23}$$

$$Y = 100\sin(\frac{1}{5}\omega t) + 10\sin(\frac{6}{5}\omega t) \tag{24}$$

respectively. Because of exceeding linear velocity ( $\Delta v > 0$ ) of a point p (on the circumference of $C_b$) due to equations (21) and (22), $C_b$ is sliding back (backward sliding) along the virtual circle $C_{a-b}$ . The traced path of point p (on the circumference of $C_b$) for this situation is shown in the left side of Fig2c.Same argument would be applicable for (23) and (24). The only difference of the last case come back to the sliding mode of $C_b$ that is forward sliding ( $\Delta v < 0$ ). The traced path of point p based on equations (23) and (24) is shown in the right side of Fig2c.

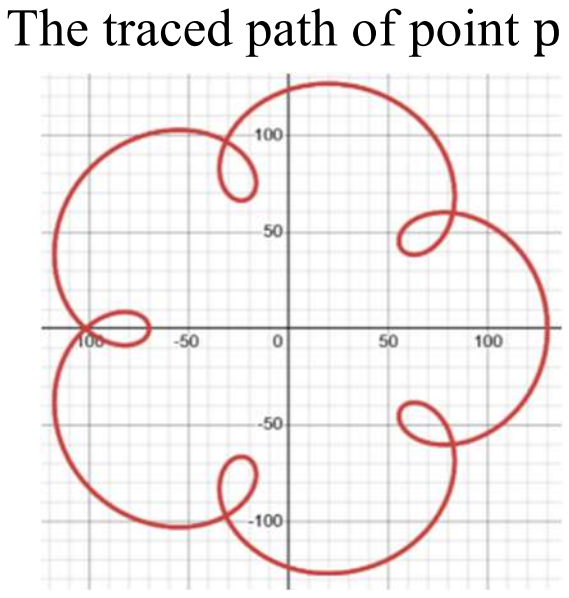

Fig2c **Left side**: Backward sliding of virtual circle $C_b$ along the virtual circle $C_{a-b}$. **Right side:** forward sliding of virtual circle $C_b$ along the virtual circle $C_{a-b}$. Both geometrical objects are plotted through VRCT by computer.

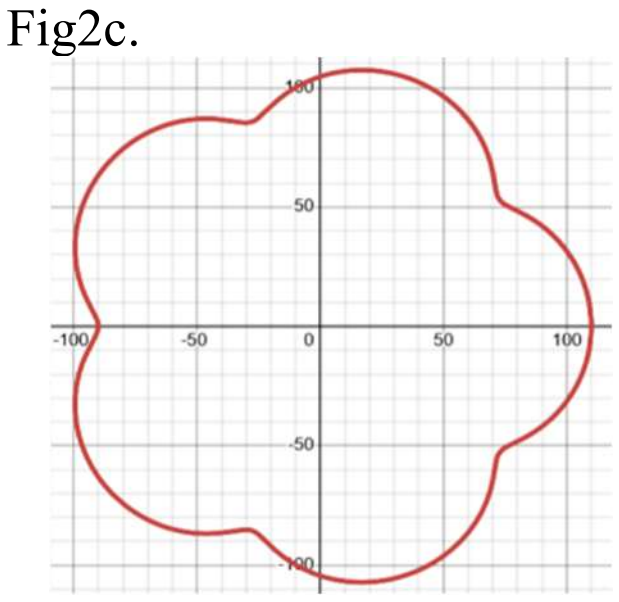

Now, we are in a stage to generalize and formulate the parametric equations of an epitrochoid on the bases of uniform combination of rolling and sliding motions through UCRSMCP. Suppose that the parameters $a$ , $b$ , $\omega$ and $\Omega$ are selected such that equation (14) is satisfied (pure rolling motion). we assume that parameter b is changed by $\Delta b$ in equations (5) and (6) while other parameters are considered to be fixed. Therefore, in new situation equations (5) and (6) are changed as follows

$$X = a\cos\Omega t + (b + \Delta b)\cos(\omega + \Omega)t \quad (25)$$

$$Y = a\sin\Omega t + (b + \Delta b)\sin(\omega + \Omega)t \quad (26)$$

From physical point of view if $\Delta b > 0$ then, linear speed of a point on the circumference of virtual circle $C_b$ is increased due to the increment of $\Delta b$ and vice versa and sliding motion would be occurred. If $\Delta S$ is the sliding arc length in one complete rotation of the circle $C_{a-b}$ , then

$$\Delta S = T_\Omega \Delta v \quad , \quad T_\Omega \text{ is time period of tablet1 (turn table)} \quad (27)$$

Therefore,

$$\Delta S = \frac{2\pi}{\Omega}[(b + \Delta b)\omega - [a + \Delta b - (b + \Delta b)]\Omega] = \frac{2\pi\omega\Delta b}{\Omega} \quad (28)$$

Thus,

$$\frac{\Delta S}{2\pi} = \frac{\omega}{\Omega}\Delta b \quad \text{or} \quad \Delta b = \frac{\Omega}{\omega}\frac{\Delta S}{2\pi} \quad (29)$$

*Equations (25) and (26) reflect the usual mathematical vision for creating centered trochoids through pure rolling motion beside an attached point to a rolling circle*! By using equations (25), (26) and (29) we can deduce the parametric equations of the traced path of a point p on the circumference of $C_b$ through combination rolling and sliding motions of circle $C_b$ along circle $C_{a-b}$. *In the other words, we have discovered an innovative and more conceptive mental vision to define centered trochoids on the bases of physical concepts for oscillatory systems beside combination of rolling and sliding motions!*

Thus, in the new situation on the bases of rolling and sliding motions equations (5) and (6) are changing as follows:

$$X = a\cos\Omega t + b\cos(\omega + \Omega)t + \frac{\Omega}{\omega}\frac{\Delta S}{2\pi}\cos(\omega + \Omega)t \quad (30)$$

$$Y = a\sin\Omega t + b\sin(\omega + \Omega)t + \frac{\Omega}{\omega}\frac{\Delta S}{2\pi}\sin(\omega + \Omega)t \quad (31)$$

It is obvious that for pure rolling motion $\Delta S$ must be zero and equations (30) and (31) would be again same as the parametric equations (5) and (6). In the other words equations (5) and (6) are especial case of (30) and (31) respectively.

We come back to calculate $\frac{\Delta S}{2\pi}$ for $\Delta b = 30 - \frac{50}{3}$ and $\Delta b = 10 - \frac{50}{3}$ (given by example2) respectively.

$$\frac{\Delta S}{2\pi} = \frac{\omega}{\Omega}\Delta b = \frac{1}{\frac{1}{5}}\left(30 - \frac{50}{3}\right) = \frac{200}{3} \text{ unit of arc length/radian} \quad (32)$$

$$\frac{\Delta S}{2\pi} = \frac{\omega}{\Omega}(\Delta b) = \frac{1}{\frac{1}{5}}\left(10 - \frac{50}{3}\right) = \frac{-100}{3} \text{ unit of arc length/radian} \quad (33)$$

Thus, for backward sliding case in example2 we have

$$X = 100\cos(\frac{1}{5}\omega t) + \frac{50}{3}\cos(\frac{6}{5}\omega t) + \frac{40}{3}\cos(\frac{6}{5}\omega t) = 100\cos(\frac{1}{5}\omega t + 30\cos(\frac{6}{5}\omega t) \quad (34)$$

$$Y = 100\sin(\frac{1}{5}\omega t) + \frac{50}{3}\sin(\frac{6}{5}\omega t) + \frac{40}{3}\sin(\frac{6}{5}\omega t) = 100\sin(\frac{1}{5}\omega t) + 30\sin(\frac{6}{5}\omega t) \quad (35)$$

For forward sliding case of example2 we have

$$X = 100\cos(\frac{1}{5}\omega t) + \frac{50}{3}\cos(\frac{6}{5}\omega t) - \frac{20}{3}\cos(\frac{6}{5}\omega t) = 100\cos(\frac{1}{5}\omega t) + 10\cos(\frac{6}{5}\omega t) \quad (36)$$

$$Y = 100\sin(\frac{1}{5}\omega t) + \frac{50}{3}\sin(\frac{6}{5}\omega t) - \frac{20}{3}\sin\left(\frac{6}{5}\omega t\right) = 100\sin(\frac{1}{5}\omega t) + 10\sin(\frac{6}{5}\omega t) \quad (37)$$

Equations (34) and (35) [also (36) and (37)] are new versions of parametric equations (19) and (20) for backward [forward] sliding modes respectively. The geometrical objects for some different values of $\frac{\Delta S}{2\pi}$ are shown in the left side of Fig2d. Note that there is restriction on decreasing $\Delta b$ so that $\Delta b$ cannot be less than $-b$. It is obvious the traced path in this situation is a circle with radius a. Fig2d shows a number of traced paths of a point p (on the circumference of a virtual circle $C_b$ that is given in example2) These geometrical objects are the results of combination rolling and sliding motions of $C_b$ along a virtual circle $C_{a-b}$ through changing parameter.

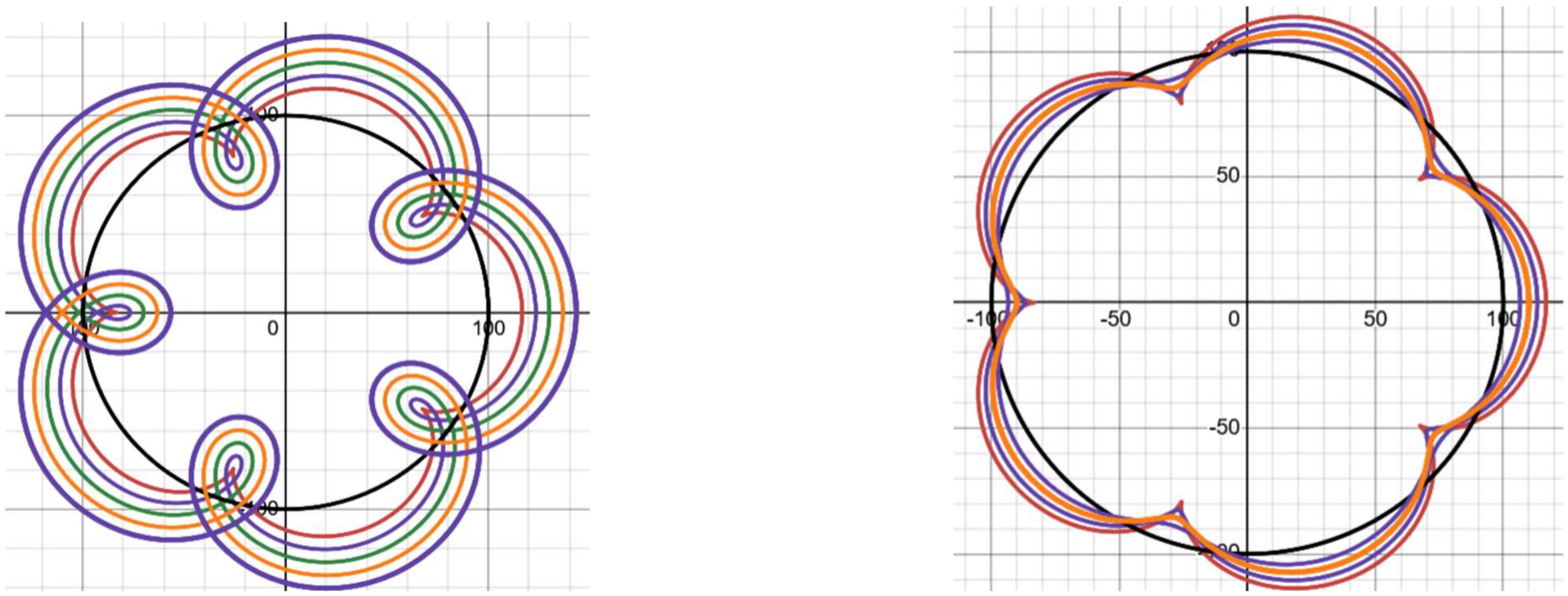

**Fig2d**. Uniform Combination of Rolling and Sliding Motions through Changing Parameter (UCRSMCP). Circles in black color are due to pure sliding motion of a circle that its radius is zero along another one that its radius is 100 units. **Left side** (Backward sliding): Red is an epicycloid due to pure rolling motion. Purple orbit is an epitrochoid that it has most backward sliding motion relative to the others. **Right side**: (Forward sliding): Epicycloid (due to pure rolling motion) is shown by red. Blue object has most forward sliding motion after the the circle.

**2C. Uniform Combination of Rolling and Sliding Motions through Changing Frequency (UCRSMCF)**

Process of combination rolling and sliding motions for a virtual circle $C_b$ along another virtual circle $C_{a-b}$ can be performed by changing angular frequency of $C_b$ or $C_{a-b}$ through a different perspective. we are doing this process through the changing angular frequency of $C_b$ while other parameters are assumed to be constant. Suppose that the parameters $a, b, \omega$ and $\Omega$ are selected such that equation (14) is satisfied for pure rolling motion. Obviously if we change $\omega \rightarrow \omega + \Delta\omega$ then (14) is violated. Thus, $(a-b)\Omega \neq b\omega$. We define

$$\Delta v = b(\omega + \Delta\omega) - (a-b)\Omega \tag{38}$$

Obviously, it is clear that $\Delta\omega > 0$ requires $\Delta v > 0$ such that operation of the system would be on backward sliding mode. Sliding arc length in one complete rotation of turntable is as follows

$$\Delta S = \frac{2\pi}{\Omega}[b(\omega + \Delta\omega) - (a-b)\Omega] \tag{39}$$

From (14) and (39) we have

$$\frac{\Delta S}{2\pi} = \frac{b\Delta\omega}{\Omega} \qquad \text{or} \qquad \Delta\omega = \frac{\Omega}{b}\frac{\Delta S}{2\pi} \tag{40}$$

By using (5), (6) and (40) the parametric equations of traced path are given by

$$X = a\cos\Omega t + b\cos\left(\omega + \Omega + \frac{\Omega}{b}\frac{\Delta S}{2\pi}\right)t \tag{41}$$

$$Y = a\sin\Omega t + b\sin\left(\omega + \Omega + \frac{\Omega}{b}\frac{\Delta S}{2\pi}\right)t \tag{42}$$

**Example 3.** On the bases of equation (14) and VRCT an epicycloid can be created with following parameters: $a = 140,\ \Omega = 1\ ,\ b = 20\ ,\ \omega = 6$. Compute $\frac{\Delta S}{2\pi}$ for $\Delta\omega = \pm1, \pm2, \pm3, \pm4, \pm5,\ \pm6$ and plot the corresponding geometrical objects.

$\frac{\Delta S}{2\pi} = \frac{b\Delta\omega}{\Omega} = 20\,[\pm1, \pm2, \ldots] = \pm20\ , \pm40\ , \pm60$ ... Therefore, from equations (41) and (42)

$$X = 140\cos t + 20\cos\left(7 + \frac{1}{20}\,(\pm20\ , \pm40\ , \pm60\ , \ldots\ )\right)t \tag{43}$$

$$Y = 140\sin t + 20\sin\left(7 + \frac{1}{20}\,(\pm20\ , \pm40\ , \pm60\ , \ldots\ )\right)t \tag{44}$$

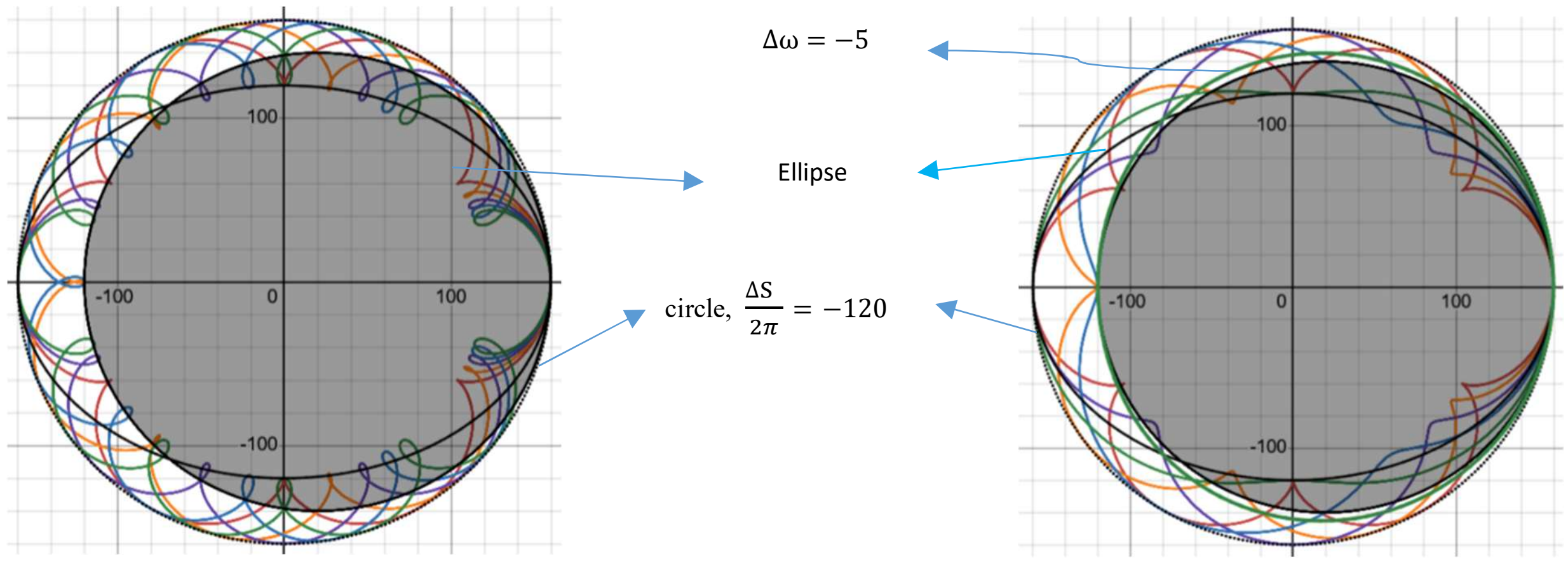

**Fig2e**. Uniform Combination of Rolling and Sliding Motions through Changing Frequency UCRSMCF. In both pictures simulation of pure rolling motion is shown by red. Shaded circle is due to $\Delta\omega = -7$ In both pictures' ellipse is due to $\Delta\omega = -8$ **Left side**: Colored geometrical objects (except red) are epitrochoids on backward sliding mode. **Right side:** Colored geometrical objects (except red) are epitrochoids on forward sliding mode. Green geometrical object near the shaded circle is an epitrochoid ( $\Delta\omega = -5$ ).

### 3A. Analytical Details for plotting a Hypocycloid Through VRCT

According to the definition, a hypocycloid is a plane curve generated by a fixed point on the circumference of a circle (with radius b) that rolls purely within a circle with radius $R > b$ .If $2\pi R = 2\pi N b$ $(N = 2,3,4,5, \dots.)$ Then, the hypocycloid includes N repeated parts in $2\pi$ radians. In order to create a hypocycloid through VRCT the rolling condition requires different angular frequencies for tablet1 and tablet2. [radius of the virtual circle $C_{a+b}$ on the turn table is $R = a + b$ Fig3a]. In this situation both virtual circles $C_b$ and $C_{a+b}$ must be rotating in the same direction as shown by Fig3a (polarization issue). Suppose that $\omega$ and $\Omega$ to be angular frequencies of tablets2 and tablet 1 respectively. On the bases (VRCT) rolling condition requires that equation (13) to be satisfied. We consider a hypocycloid with $n_h$ repeated parts (in one complete rotational motion of turntable) then, obviously radius b is given by

$$b = \frac{a}{n_h - 1} \qquad n_h = 2,3,4, \dots. \tag{45}$$

By combining formulas (45) and (13) we have

$$n_h = \frac{\omega}{\Omega} \tag{46}$$

Equation (45) requires that $n_h \geq 2$ since b can not be greater than a .Case $n_h = 2$ leads us to interesting results. In this situation from (45) and (46) we have $b = a$ and $\omega = 2\Omega$. This means that diameter of the virtual circle $C_b$ is equal to the radius of virtual circle $C_{a+b}$ while the angular frequency of tablet2 is twice the angular frequency of the tablet1. In the next section we will show that for $n_h = 2$ , created object through VRCT would be a line segment (diameter of virtual circle $C_{a+b}$ as shown by Fig3b). On the other hand, in the framework of the requirements $n_h = 2$ and $\omega = 2\Omega$ it is possible to visualize other different possibilities. If $b \neq a$ then, we are going to show that outputs of this situation would be co- centered ellipses in forward and backward sliding modes!

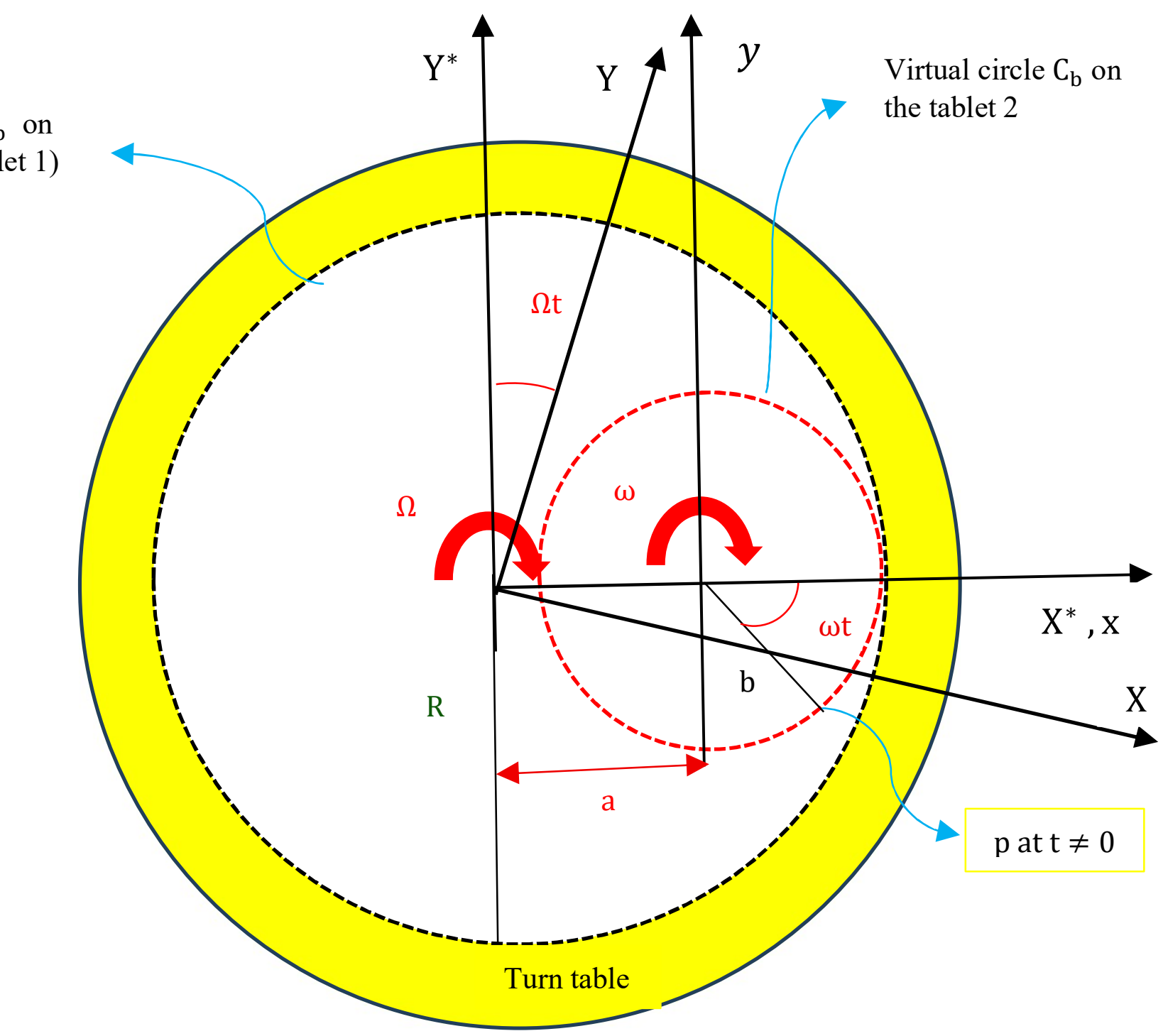

**Fig3a** Virtual circles on the tablet 1 and tablet 2 for plotting hypocycloid through the VRCT. Virtual circles are shown by dashed lines. XY coordinate system is fixed on the turn table and $X^*Y^*$ is a stationary coordinate system. This figure shows mathematical simulation for rolling a circle $C_b$ along another virtual circle $C_{a+b}$ .

**3B. Creating Ellipse through VRCT** [Uniform Combination of Rolling and Sliding Motions through Changing Parameter (UCRSMCP)]

In this section we are going to extend a surprising mathematical method for creating ellipse based on uniform combination of rolling and sliding motions. This interesting beautiful concept can be described as combination of two co- polarized periodic motions with two different angular frequences! Although creation an ellipse through the combination of two orthogonal simple harmonic oscillations with same frequencies is a standard known method in scientific references of physics and mathematics, but we are going to introduce a different innovative perspective for creating ellipse on the bases of combination of two periodic motions with different frequencies that can be also interpreted as combination of rolling and sliding motions. In the Fig3a $X^*Y^*$coordinate system is considered to be a stationary laboratory system while XY coordinate system rotates with angular frequency $\Omega$. We assume that at the

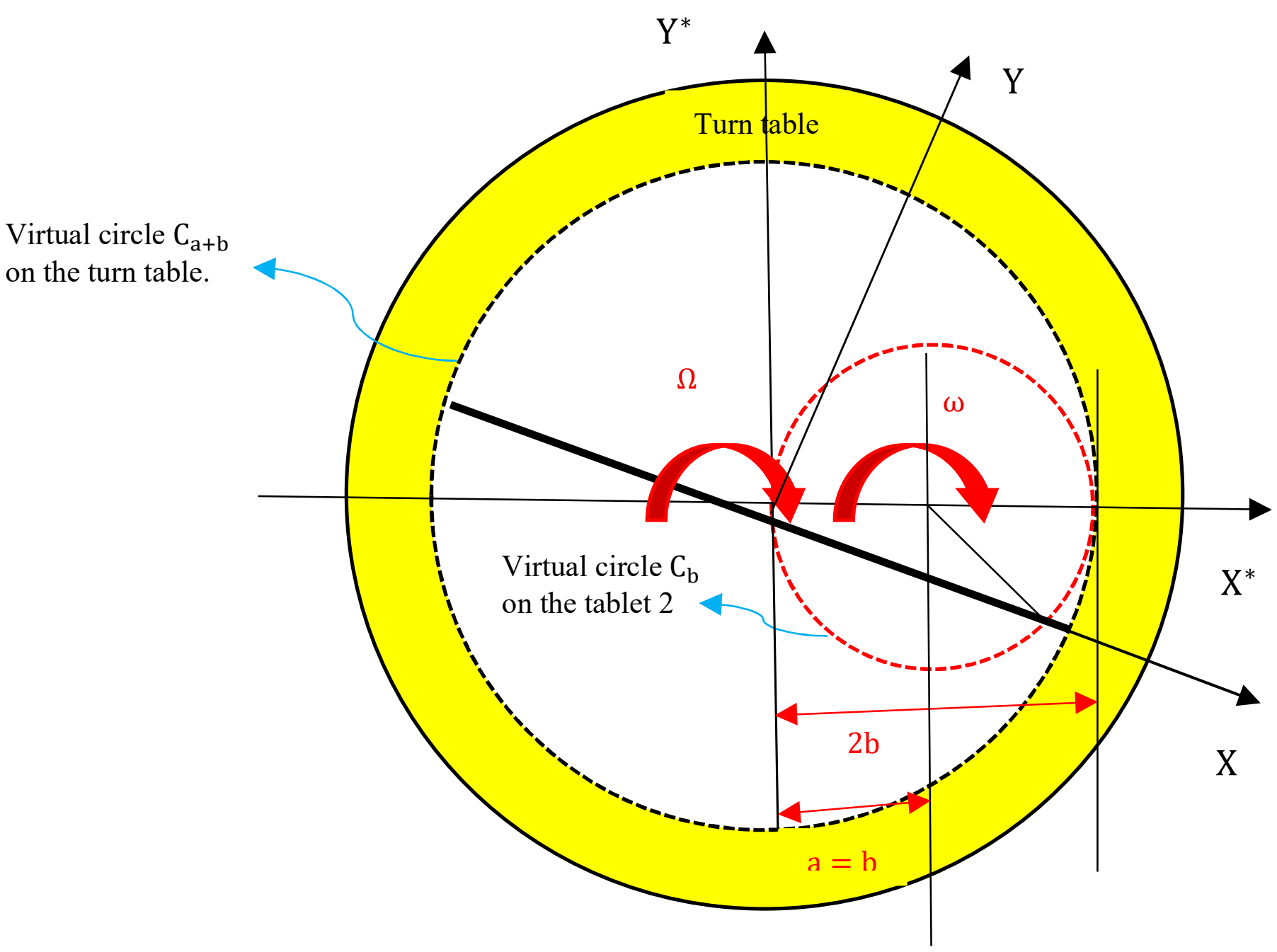

**Fig3b**Configuration of virtual circles on rotating tablets when $n_h = 2$ , ( a = b) The bold line segment shows the path of the pen on turn table. Pen dose a simple harmonic oscillation along the line segment in the rotating cartesian coordinate system XY that is fixed on the turn table. Note that in this situation ($\omega = 2\Omega$).

moment $t = 0$, the corresponding axis of XY and $X^*Y^*$ coordinate systems are coincided. On the bases of equation (46) $n_h = 2$ requires that $\omega = 2\Omega$. Thus, by using Fig3a, the position of pen in $X^*Y^*$ coordinate system at the moment $t \neq 0$ is given by

$$X^* = a + b\cos 2\Omega t \tag{47}$$

$$Y^* = -b \sin 2\Omega t \tag{48}$$

The coordinate transformations (3) and (4) require that

$$X = X^* \cos\Omega t - Y^* \sin\Omega t = [a + b\cos 2\Omega t]\cos\Omega t + b \sin 2\Omega t \sin\Omega t \tag{49}$$

$$Y = X^* \sin\Omega t + Y^* \cos\Omega t = [a + b\cos 2\Omega t]\sin\Omega t - b \sin 2\Omega t \cos\Omega t \tag{50}$$

Therefore,

$$X = (a + b) \cos\Omega t \tag{51}$$

$$Y = (a - b) \sin\Omega t \tag{52}$$

Equations (51) and (52) are clearly parametric equations of an ellipse. Requirement $a = b$ for equations (51) and (52) leads us to $Y = 0$ and $X = 2b \cos\Omega t$. Therefore, in this situation, pen does simple harmonic oscillations on the rotating X axis and a line segment (special case of an ellipse) will be created. We encounter an interesting result here: As we stated before, $\omega = 2\Omega$ requires that $a = b$ for pure rolling motion of $C_b$. Therefore, the line segment (special case of an ellipse) is a geometrical object that is created through the pure rolling $C_b$ inside $C_{a+b}$. On the other hand, when $\omega = 2\Omega$ but $a \neq b$ on the bases of equations (51) and (52) the created geometrical object would be an ellipse. In this case the motion of $C_b$ along $C_{a+b}$ is not a pure rolling motion. Another especial case of an ellipse is a circle that is created through a pure sliding motion of virtual circle $C_b$ inside $C_{a+b}$. Thus, line segment is only an especial case of an ellipse that is created through a pure rolling motion while all the other shapes of ellipses (except circle) are created through uniform combinations of rolling and sliding motions of different virtual circles $C_b$ inside other virtual circles $C_{a+b}$. By using (51) and (52) it is clear that

$$\frac{X^2}{A^2} + \frac{Y^2}{B^2} = 1 \qquad A = a + b \ , \qquad B = a - b \tag{53}$$

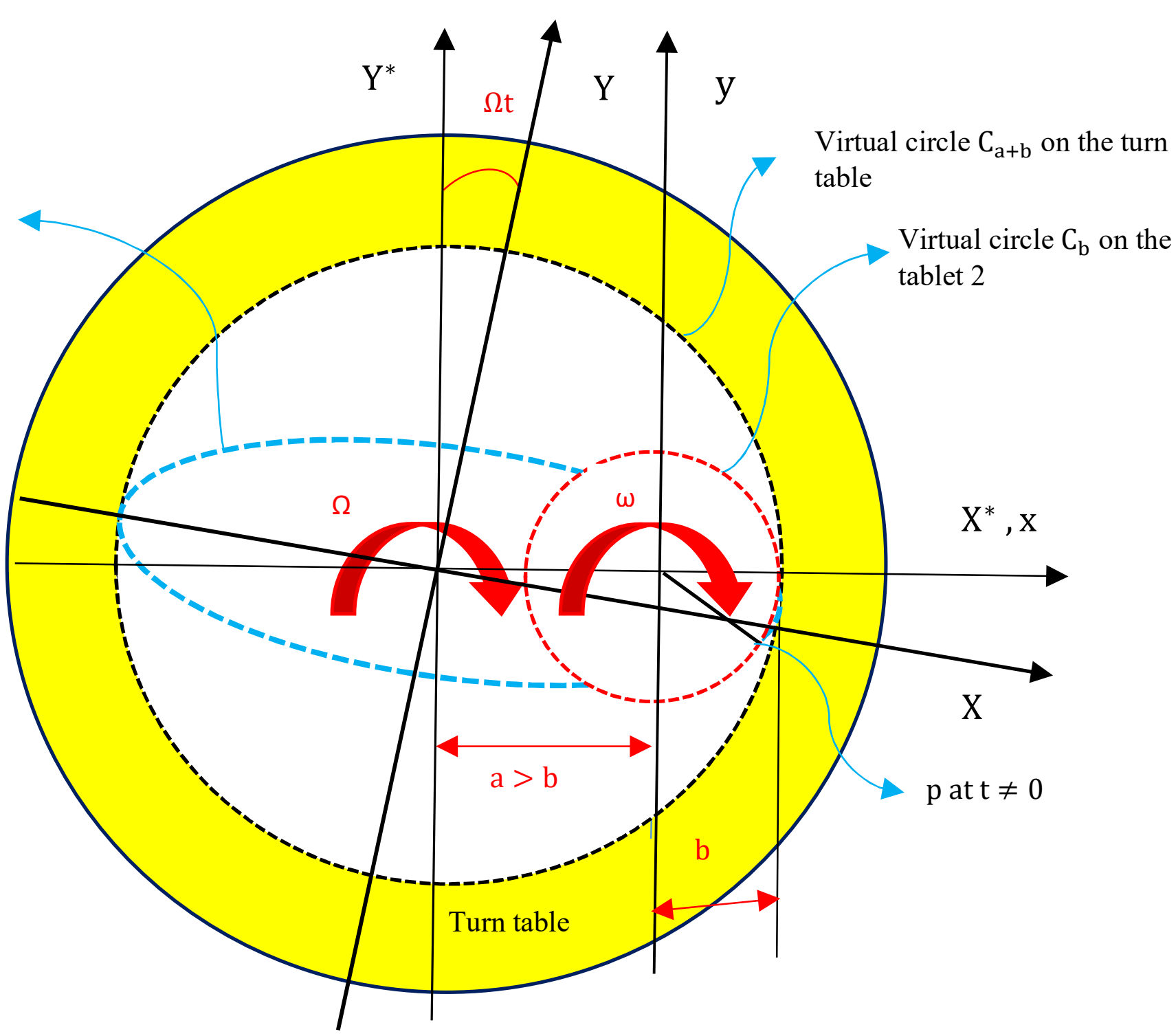

**Fig3c.** An ellipse (bold dashed line) is created through the combination of rolling and sliding motions in forward sliding mode. Other thin dashed lines are virtual circles (black) on the tablet 1, (red) on tablet 2. XY (rotating system) is fixed on the turn table. The stared coordinate system is a stationary laboratory coordinate system. This figure shows mathematical simulation of co-polarized situation for combination of rolling and sliding motions (of circle $C_b$) along circle $C_{a+b}$. ($\omega = 2\Omega$)

On the bases of combination rolling and sliding motions, it is clear that creating different ellipses can be done only through UCRSMCP by changing radius of $C_b$ or $C_{a+b}$ .Here we are going to do this subject by changing radius of $C_b$ while radius of $C_{a+b}$ is assumed to be fixed. In order to creating different co- centered ellipses, there is no selective possibility on changing commensurable angular frequencies ω and Ω because $\frac{\omega}{\Omega} = 2$. In process of creating co-centered ellipses through VRCT, creation a line segment (especial case of ellipse) requires pure rolling motion by virtual circle $C_b$ along $C_{a+b}$ . This is so, when $b = a$ as shown by Fig3b. In this situation linear speed of a point on circumference of $C_b$ is same as the linear speed of a point on the circumference of $C_{a+b}$ . Therefore, when $b = a$ there is no sliding motion in the simulation process. But in general case when $b = b + \Delta b$ , pure rolling motion would be violated. Thus, in such a situation motion of $C_b$ along $C_{a+b}$ would be a definite combination of rolling and sliding motions. Therefore, we take $\Delta b > 0$ and define $\Delta v$ as follows:

$$\Delta v = (b \pm \Delta b)\omega - (a \mp \Delta b + b \pm \Delta b)\Omega \quad ,(b = a) \tag{54}$$

We take $B' = b \pm \Delta b$ as radius of new virtual circle $C_b$ and $A' = a \mp \Delta b$ as distance between the centers of $C_b$ and $C_{a+b}$ in new configuration. As was stated before creating ellipse through VRCT requires $\omega = 2\Omega$ . Then (54) takes the following simple form:

$$\Delta v = -(A' - B')\Omega \tag{55}$$

Therefore, $A' > B'$ leads us to create ellipses on the forward sliding mode while $A' < B'$ leads us to the backward sliding mode for creating ellipse. Thus, sliding arc $\Delta s$ by $C_b$ along $C_{a+b}$ in one complete rotation of turn table is given by

$$\Delta s = (\Delta v)T_{\Omega} \,, \ [T_{\Omega} = \frac{2\pi}{\Omega} \text{ is time period of one complete rotation of turn table}] \tag{56}$$

Through the combination of equations (55) and (56) we have

$$\frac{\Delta s}{2\pi} = -(A' - B') \tag{57}$$

For a definite value of $|\Delta s|$, an ellipse that is created on forward sliding mode is not distinguishable from an ellipse that is created on backward sliding mode! This important point is reflected in Fig3d.

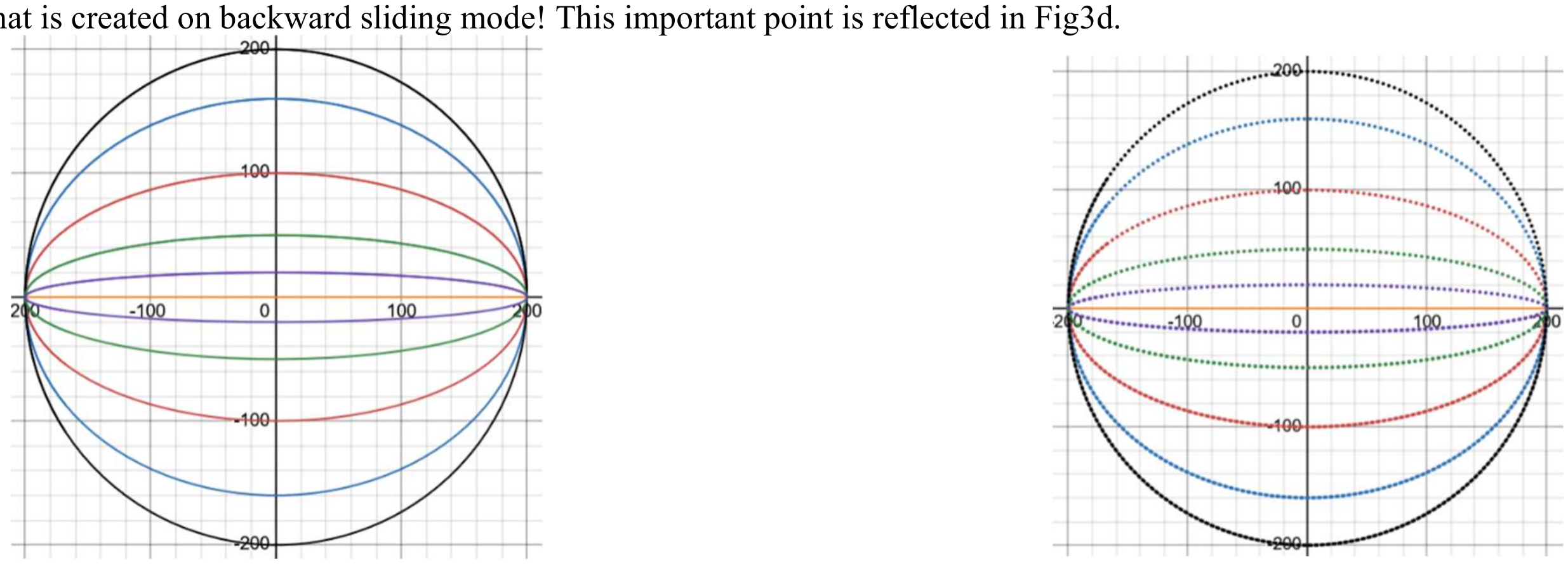

**Fig3d**. **Left side**: $A' > B'$ (**Forward sliding**). Circle (black) $A' = 200\,, B' = 0\,, \frac{\Delta s}{2\pi} = -200$ (pure sliding case). Line segment (orange) $A' = 100\,, B' = 100\,, \frac{\Delta s}{2\pi} = 0$ (pure rolling case). Ellipse (blue) $A' = 180\,, B' = 20\,, \frac{\Delta s}{2\pi} = -160$. Ellipse (red) $A' = 150\,, B' = 50,\ \frac{\Delta s}{2\pi} = -100$). Ellipse (green) $A' = 125\,, B' = 75, \frac{\Delta s}{2\pi} = -50$ Ellipse (purple) $A' = 110\,, B' = 90\,, \frac{\Delta s}{2\pi} = -20$ **Right side**: $A' < B'$ (**Backward sliding**). Circle (black) $A' = 0\,, B' = 200\,, \frac{\Delta s}{2\pi} = 200$ (pure sliding case). Line segment (orange) $A' = 100\,, B' = 100\,, \frac{\Delta s}{2\pi} = 0$ (pure rolling case). Ellipse (blue) $A' = 20\,, B' = 180,\ \frac{\Delta s}{2\pi} = 160$. Ellipse (red) $A' = 50\,, B' = 150\,, \frac{\Delta s}{2\pi} = 100$. Ellipse (green) $A' = 75\,, B' = 125\,, \frac{\Delta s}{2\pi} = 50$. Ellipse (purple) $A' = 90\,, B' = 110\,, \frac{\Delta s}{2\pi} = 20$

Analysis has been done in this section is considered for creating ellipse on the bases of combination rolling and sliding motions. This is an innovative method for plotting co-centered ellipses and leads us to the following theorem:

**Theorem 2**. Assume that:

1. $P(a + b, 0)$ is a definite point on the circumference of a virtual circle $C_b$ as shown in Fig3a
2. Centers of virtual circles $C_b$ (with radius b) and $C_{a+b}$ (with radius $a + b$) are fixed on the axis $X^*$(axis of stationary coordinate system $X^*Y^*$ )
3. Tablet 1 and tablet 2 (along with virtual circles $C_{a+b}$ and $C_b$) are rotating in the same direction respectively.
4. XY coordinate system rotates in- phase along with turn table (tablet 1) with angular frequency Ω.
5. Virtual circle $C_b$ is rotating along with tablet 2 with angular frequency ω.

Then, if $\omega = 2\Omega$ and $A' \neq B'$ the path is swept by point p on the plane of XY due to rotation of tablets of MS machine would be an ellipse [Fig 3c]. If $A' = B'$ the point p does simple harmonic oscillations along the rotating axis X.
If $A' = 0$ and $\omega \neq \Omega$, co-centered circles will be created due to changing $B'$. Also, due to $A' \neq 0$ and $B' = 0$ we can again create co-centered circles on pure forward sliding mode through the changing parameter $A'$. In these two last cases the function of VRCT would be same as the function of a simple math compass.

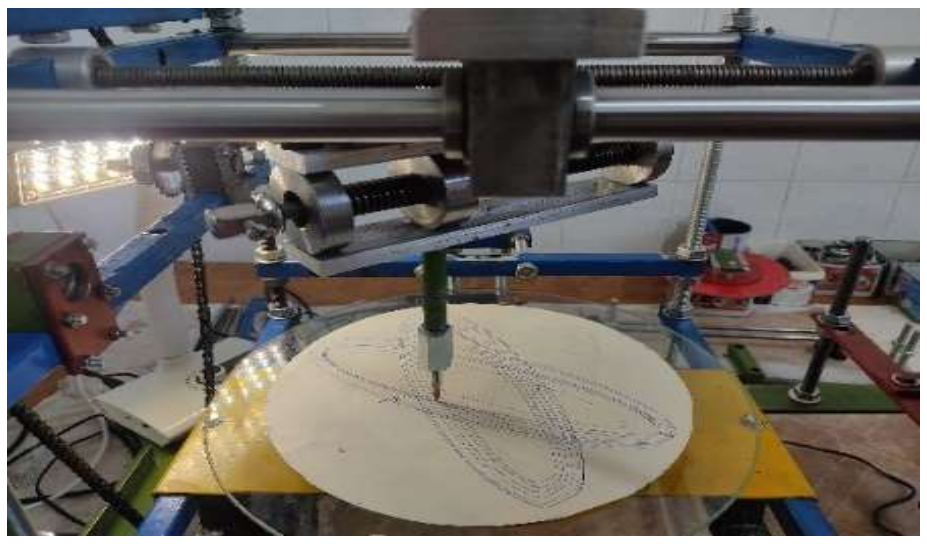

**Picture 2 Left side**: Co-centered ellipses on the turn table of MS machine.

**Right side**: A line segment as an especial case of ellipse that is created through VRCT ($n_h = 2$ , $A' = B'$) . This situation simulates pure rolling motion of circle $C_b$ inside another circle $C_{a+b}$ Fig3b.

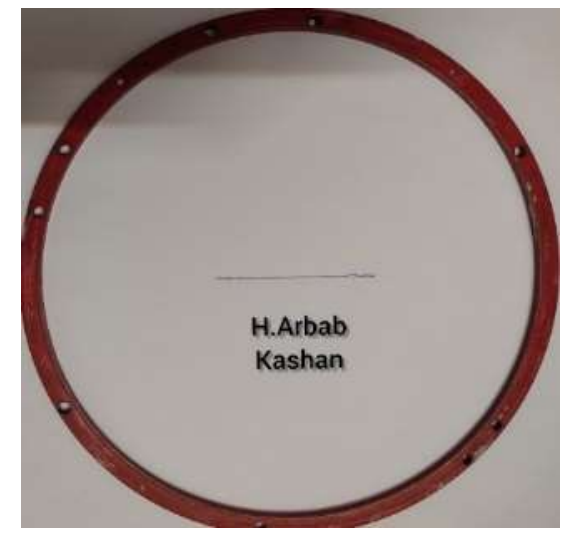

### 3C. Creating Hypocycloid through Combination of Rolling and Sliding Motions

On the bases of analysis was done in section 3A, plotting hypotrochoids require that $n_h > 2$ and polarization issue requires tablets rotate in same direction. Therefore, creating a hypocycloid requires that equation (13) to be satisfied.
**Example 4.** Assume that $a = 100$ , $\omega = 1$ and we want to try create a hypocycloid with five repeated parts in one complete rotation of tablet 1. Compute b, $\Omega$ and determine parametric equations on the bases VRCT then plot the hypocycloid. In the first step equation (45) reqires that $b = \frac{100}{5-1} = 25$ . Also, equation (13) requiers $\Omega = \frac{1}{5}$ . Thus,by using equations (9) and (10) beside the last resuts for $\Omega$ and b we have

$$X = 100\cos\frac{1}{5}t + 25\cos\frac{4}{5}t \tag{58}$$

$$Y = 100\sin\frac{1}{5}t - 25\sin\frac{4}{5}t \tag{59}$$

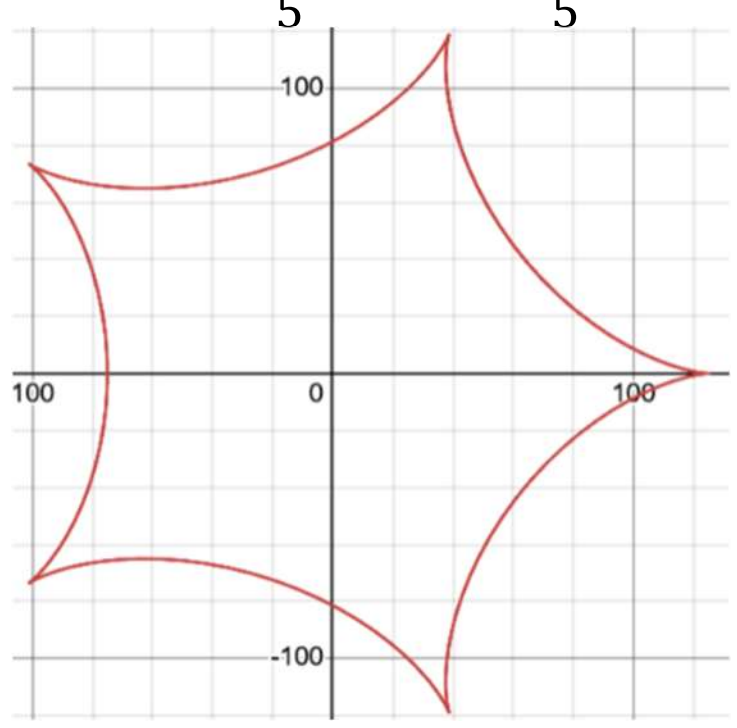

**Fig4a. Left side**: A hypocycloid with five repeated parts in $2\pi$ radians rotation of turn table **Right side**: An epicycloid with three repeated parts. It is due to replacement – sign in equation (59) with + sign. This subject reflects the important role of polarization issue in crating centered trochoids.

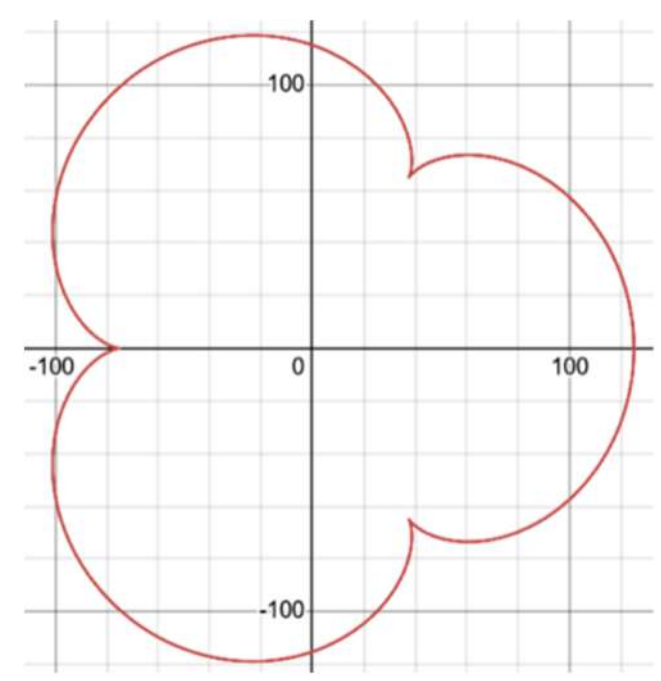

### 3D. Creating Hypotrochoid [Uniform Combination of Rolling and Sliding Motions by Changing Parameter (UCRSMCP)]

Suppose that the parameters a , b , $\omega$ and $\Omega$ are selected such that equation (13) is satisfied. Now we assume that parameter b is changed by $\Delta b$ while other parameters are fixed. Therefore, in the new configuration equations (9) and (10) are changed as follows

$$X = a\cos\Omega t + (b + \Delta b)\cos(\omega - \Omega)t \tag{60}$$

$$Y = a\sin\Omega t - (b + \Delta b)\sin(\omega - \Omega)t \tag{61}$$

By using $\Delta v = (b + \Delta b)\omega - (a - \Delta b + b + \Delta b)\Omega$ along with equations (13), (27) we have

$$\Delta v = \omega\Delta b \tag{62}$$

$$\frac{\Delta S}{2\pi} = \frac{\omega\Delta b}{\Omega} \tag{63}$$

By using equations (60), (61) and 63) we have

$$X = a\cos\Omega t + b\cos(\omega - \Omega)t + \frac{\Omega\Delta S}{2\pi\omega}\cos(\omega - \Omega)t \tag{64}$$

$$Y = a\sin\Omega t - b\sin(\omega - \Omega)t - \frac{\Omega\Delta S}{2\pi\omega}\sin(\omega - \Omega)t \tag{65}$$

*Equations (64) and (65) are new versions of equations (9) and (10) on the bases of usual mathematical deduction of hypotrochoids with the help of an attached point to a pure rolling circle.*

**Example 5.** Assume that $a = 100, \omega = 1, \Omega = \frac{1}{5}, b = 25$ if $\Delta b = 25$ determine mode of sliding motion and parametric equations on the bases VRCT. Plot graph of equations under consideration. Do the same requirments when $\Delta b = -10$ .

If $\Delta b = 25$ then, by using (63) we have $\frac{\Delta S}{2\pi} = 125$ (unit of arc length per radian). Thus, (58) and (59) convert to a new version that describe a hypotrochoid as combination of rolling and sliding motions in backward sliding mode.

$$X = 100\cos\frac{1}{5}t + 25\cos\frac{4}{5}t + 25\ \cos\frac{4}{5}t\ = 100\cos\frac{1}{5}t + 50\cos\frac{4}{5}t \qquad (66)$$

$$Y = 100\cos\frac{1}{5}t - 25\cos\frac{4}{5}t - 25\ \sin\frac{4}{5}t = 100\cos\frac{1}{5}t - 50\cos\frac{4}{5}t \qquad (67)$$

if $\Delta b = -10$ then, (63) requires $\frac{\Delta S}{2\pi} = -50$ (unit of arc length per radian). Thus, (58) and (59) convert to a new version that describes a hypotrochoid as combination of rolling and sliding motions in forward sliding mode.

$$X = 100\cos\frac{1}{5}t + 25\cos\frac{4}{5}t - 10\ \cos\frac{4}{5}t\ = 100\cos\frac{1}{5}t + 15\cos\frac{4}{5}t \qquad (68)$$

$$Y = 100\cos\frac{1}{5}t - 25\cos\frac{4}{5}t + 10\ \sin\frac{4}{5}t = 100\cos\frac{1}{5}t - 15\cos\frac{4}{5}t \qquad (69)$$

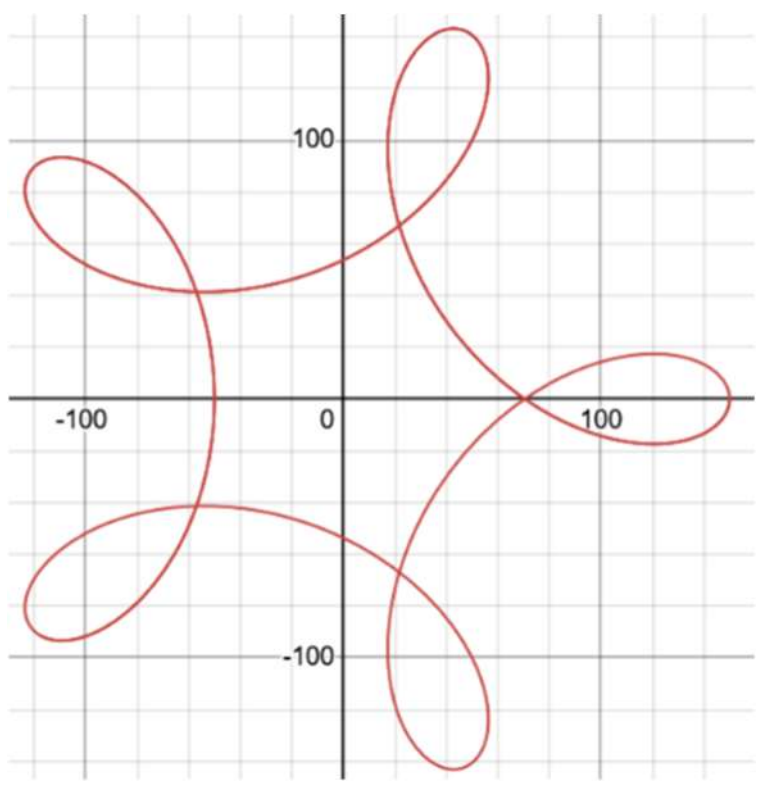

**Fig4b**. **Left side**: A hypotrochoid [equations (66) and (67)] in backward sliding mode.

($\frac{\Delta S}{2\pi} = 120$ unit of arc length per radian).

**Right side**: A hypotrochoid [equations (68) and (69)] in forward sliding mode.

($\frac{\Delta S}{2\pi} = -50$ unit of arc length per radian).

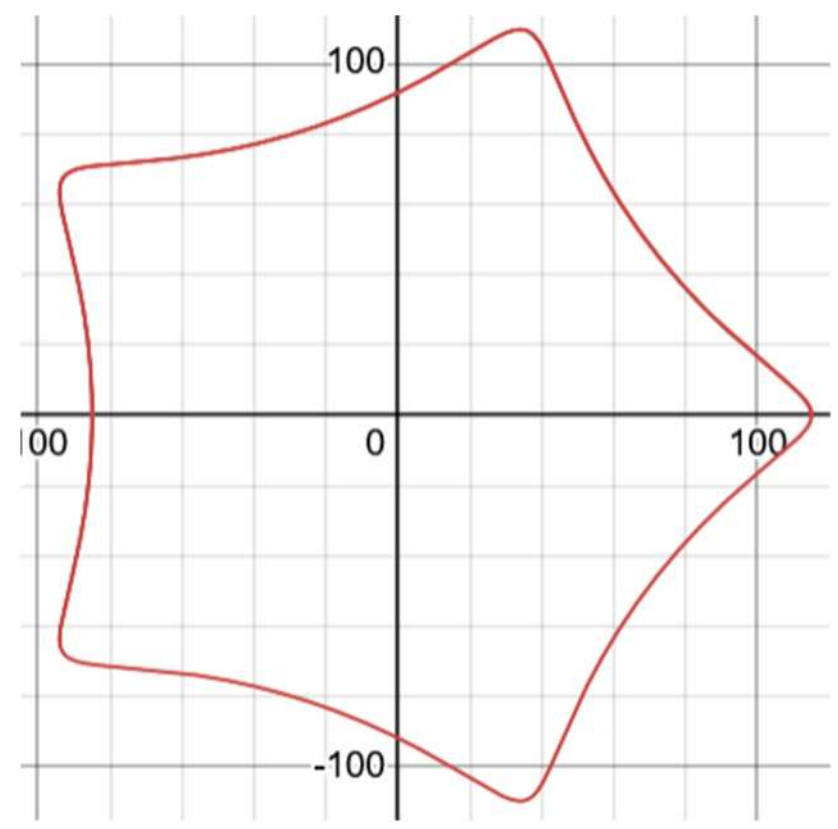

A few numbers of hypotrochoids for different values of $\frac{\Delta S}{2\pi} = 50, 75, 100, 125, 150$ (backward sliding mode) and $\frac{\Delta S}{2\pi} = -50, -75, -100, -125$ (forward sliding mode) are shown by Fig4c.

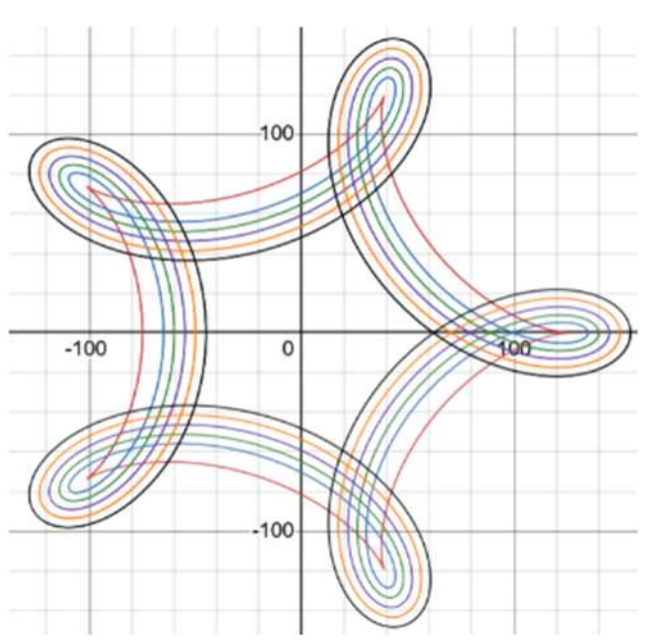

**Fig4c**. Uniform Combination of Rolling and Sliding Motions through Changing Parameter (UCRSMCP).

**Left side**: Backward sliding mode. Hypocycloid (red) $\frac{\Delta S}{2\pi} = 0$, (blue $\frac{\Delta S}{2\pi} = 50$) (green $\frac{\Delta S}{2\pi} = 75$) (purple $\frac{\Delta S}{2\pi} = 100$) (orange $\frac{\Delta S}{2\pi} = 125$) (black $\frac{\Delta S}{2\pi} = 150$) **Right side**: Forward sliding mode. Hypocycloid (red $\frac{\Delta S}{2\pi} = 0$) (blue $\frac{\Delta S}{2\pi} = -50$ ) (green $\frac{\Delta S}{2\pi} = -75$ ) (purple $\frac{\Delta S}{2\pi} = -100$) (circle in black $\frac{\Delta S}{2\pi} = -125$ )

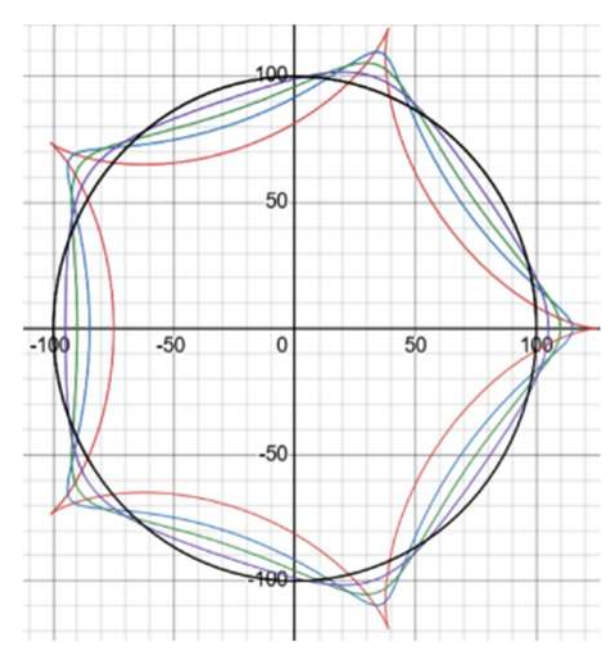

**3E. Creating Hypotrochoid [Uniform Combination of Rolling and Sliding Motions by Changing Frequency (UCRSMCF)]**

Another method for creating hypotrochoids is Uniform Combination of Rolling and Sliding Motions through Changing Frequency (UCRSMCF). In this case (13) would be violated only through changing frequencies ω or Ω while other parameters are fixed. Here we assume $\omega \to \omega + \Delta\omega$ while $\Omega, b, a$ are fixed. Therefore, $\Delta v = b\Delta\omega$ and $\Delta s = \frac{2\pi}{\Omega} b\Delta\omega$. Thus,

$$\Delta\omega = \frac{\Omega\Delta s}{2\pi b} \qquad (70)$$

In the new situation (9), (10) covert to the following equations:

$$X = a\cos\Omega t + b\cos(\omega + \Delta\omega - \Omega)t = a\cos\Omega t + b\cos\left(\omega - \Omega + \frac{\Omega\Delta s}{2\pi b}\right)t \qquad (71)$$

$$Y = a\sin\Omega t\ - b\sin(\omega + \Delta\omega - \Omega)t = a\sin\Omega t\ - b\sin\left(\omega - \Omega + \frac{\Omega\Delta s}{2\pi b}\right)t \qquad (72)$$

Thus, we obtain simply the parametric equations of hypotrochoids on the bases UCRSMCF.

**Example 6** On the bases of equation (13) beside using VRCT an hypocycloid can be created with following parameters: $a = 140, \Omega = 1$ , $b = 20$ , $\omega = 8$. Compute $\frac{\Delta S}{2\pi}$ for $\Delta\omega = \pm 1, \pm 2, \pm 3, \pm 4, \pm 5, \pm 6 \ \pm 7$ in several steps and plot the corresponding geometrical objects.

$\frac{\Delta S}{2\pi} = \frac{b\Delta\omega}{\Omega} = 20\,[\ \pm 1, \pm 2, \ldots] = \ \pm 20\ , \pm 40\ , \pm 60\ , \ldots, \pm 140$. Therefore, from equations (71) and (72) we have

$$X = 140\cos t + 20\cos\left(7\ + \frac{1}{20}\ (\pm 20\ \ , \pm 40\ , \pm 60\ , \ldots\ )\right) t \tag{73}$$

$$Y = 140\sin t\ - 20\sin\left(7\ + \frac{1}{20}\ (\pm 20\ \ , \pm 40\ , \pm 60\ , \ldots\ )\right) t \tag{74}$$

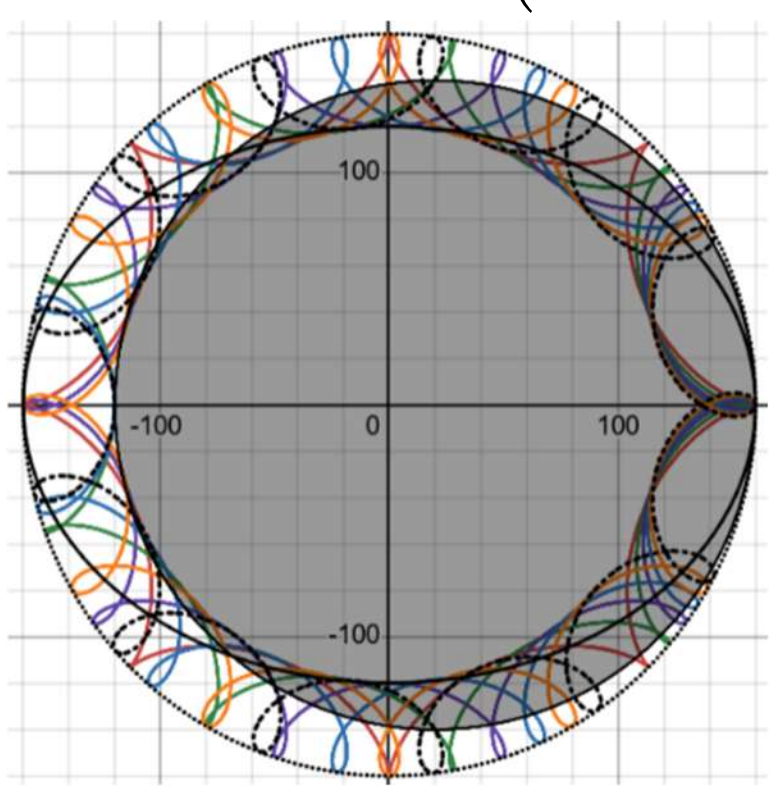

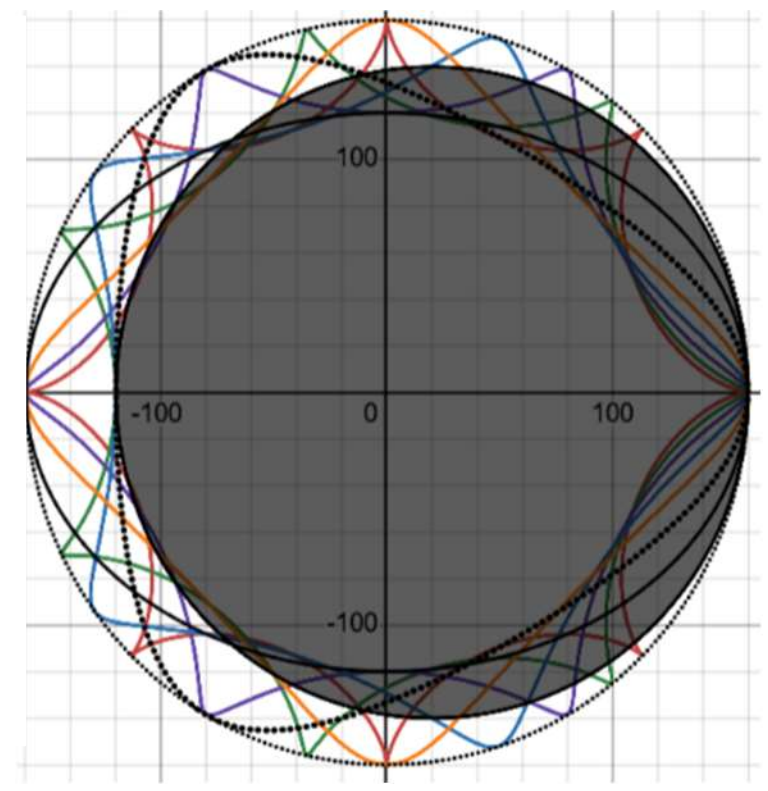

**Fig4d.** Uniform Combination of Rolling and Sliding Motions through Changing Frequency UCRSMCF. In both pictures hypocycloid is shown by red ( $\frac{\Delta S}{2\pi} = 0$ ) **Left side**: backward sliding (green, $\frac{\Delta S}{2\pi} = 20$ ) (Purple, $\frac{\Delta S}{2\pi} = 40$ ) (blue, $\frac{\Delta S}{2\pi} = 60$ ) (orange, $\frac{\Delta S}{2\pi} = 80$ ) (Dotted black, , $\frac{\Delta S}{2\pi} = 100$ **Right side**: (green, $\frac{\Delta S}{2\pi} = -20$) (Purple, $\frac{\Delta S}{2\pi} = -40$ ) (blue, $\frac{\Delta S}{2\pi} = -60$ ) (orange, $\frac{\Delta S}{2\pi} = -80$) (Dot black object, $\frac{\Delta S}{2\pi} = -100$) (ellipses, outside black circles and shaded circles in both pictures are those given in Fig2e)

A few numbers of images that are created by MS machine through VRCT:

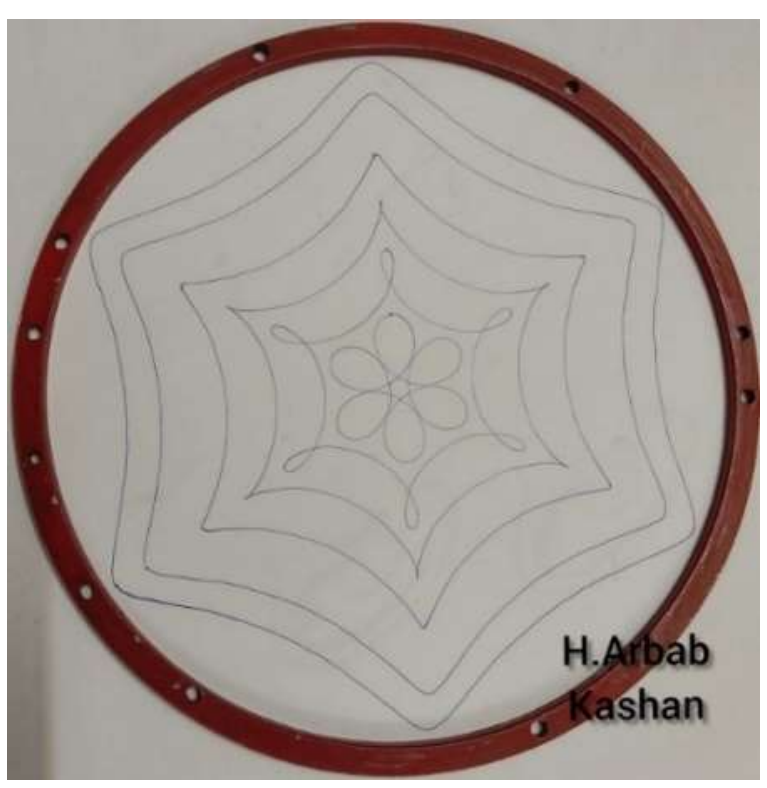

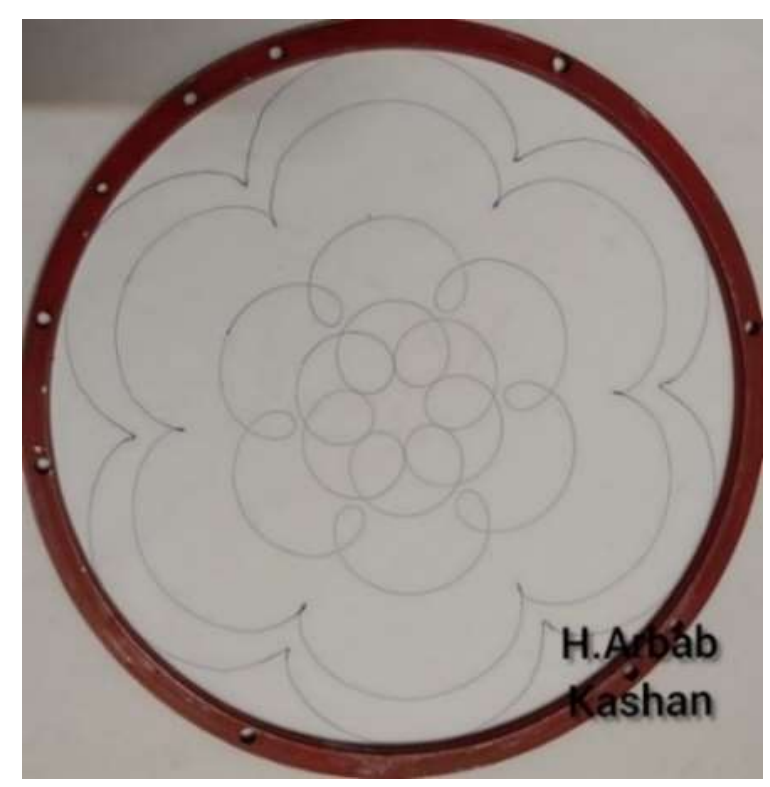

**Picture 3.** All the geometrical objects in left and right side are paths of a definite point p (on the circumference of a virtual circle $C_b$) that is rolling and sliding along another virtual circle $C_{a\pm b}$) respectively. Centered trochoids in these pictures are created by MS machine through VRCT. **Left side**: Hypotrochoids. **Right side** Epitrochoids

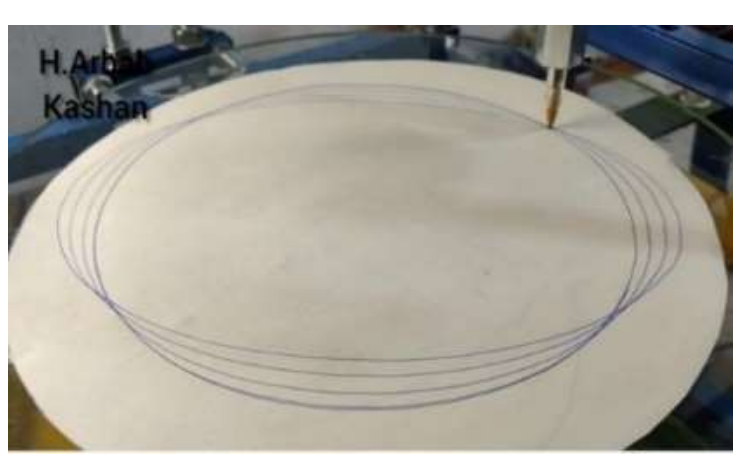

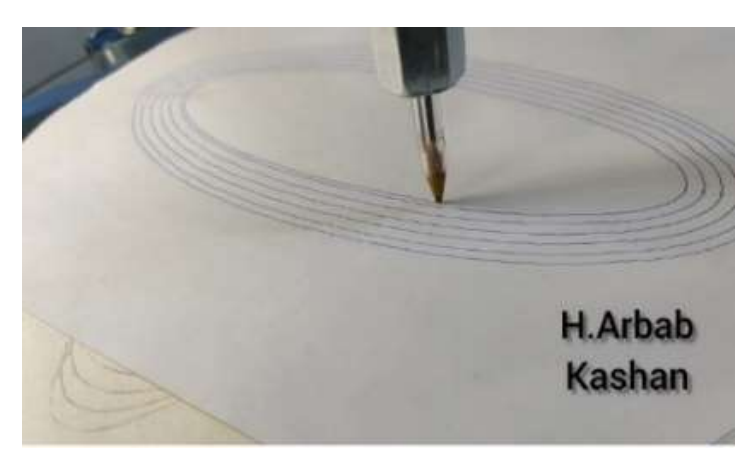

**Picture 4** Co-centered ellipses are created by MS machine.

## 4. Simulating Rolling and Sliding Motions of a Circle Along Straight Line

The practical implementation of uniform rolling and sliding motions simultaneously for a circle that is moving along straight line is not as simple as creating real rolling and sliding motions for a circle along another circle. Instead, we can implement a real rolling motion beside virtual sliding technique to simulate combination of rolling and sliding motions for a circle along straight line! In this technique, the sliding motion doesn't have reality because it is done in an imaginary fashion. In order to simulate such a situation, we use two equiaxed circles so that one of circles has a pure rolling motion along straight line but other circle rotates in phase with the pure rolling one. To ensure pure rolling motion for one of these circles along straight line, we use a gear (instead of a smooth circle) and a linear gear (as a straight line). To understand virtual sliding a circle along a straight line how must be created, consider Fig5

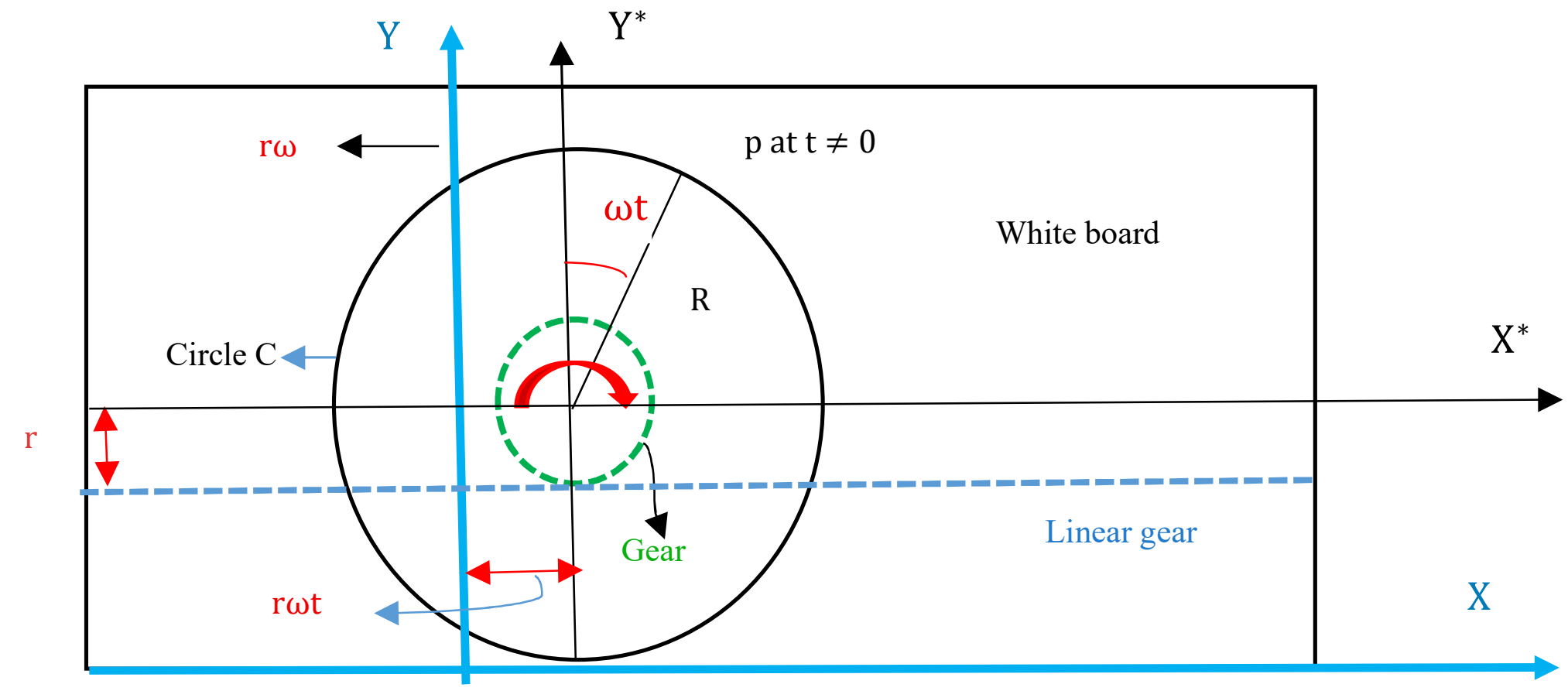

**Fig 5.** Simulating rolling and sliding motions virtually on a straight line ( X axis). Motion of gear (green dashed circle) on a linear gear (blue dashed line) is pure rolling without sliding. Circle C and gear are co- axial and they have fixed position in the laboratory coordinate system $X^*Y^*$. They are rotating with same angular frequencies ω. Linear gear is fixed in translating coordinate system XY. Motion of circle C on the X axis is combination of rolling and sliding!

Assume that a pencil is attached to the circumference of circle C at point p such that it is perpendicular to the XY plane. Then the path is swept by point p on the plane of XY would be a trochoid on backward sliding mode ($r < R$).This is so because $\Delta v = R\omega - r\omega > 0$ . On the basis of functional operation of the Virtual Sliding Simulator (VSS) that is shown by picture 5, if $r > R$ then, the traced path would be a trochoid on forward sliding mode. If $r = R$ the traced path would be a cycloid (pure rolling motion). With the help of Fig5 parametric equations of traced paths by point p are given as follows:

$$X = r\omega t + R\sin\omega t \quad (75)$$

$$Y = R + R\cos\omega t \quad (76)$$

From the fact that sliding motion of circle C is due to the first term of equation (75) then, we can define a speed $V = r\omega$ for translational motion of XY coordinate system and rewrite (75) and (76) as follows:

$$X = Vt + R\sin\omega t \quad (77)$$

$$Y = R + R\cos\omega t \quad (78)$$

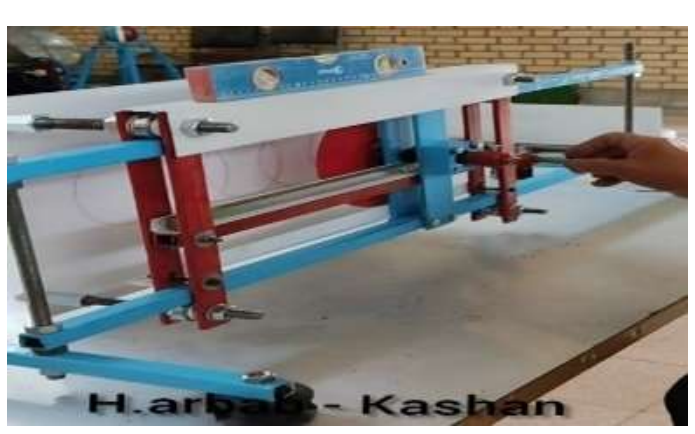

**Picture 5 Left side**: shows VSS for combination of rolling and sliding motions of a circle along straight line. In this picture there is a tablet (red) plays the role of circle C in Fig5. Trochoids are plotted on a moving white board along $X^*$. White board is fixed in XY coordinate system. **Right side**: shows a picture of created trochoid on the backward sliding mode through VSS.

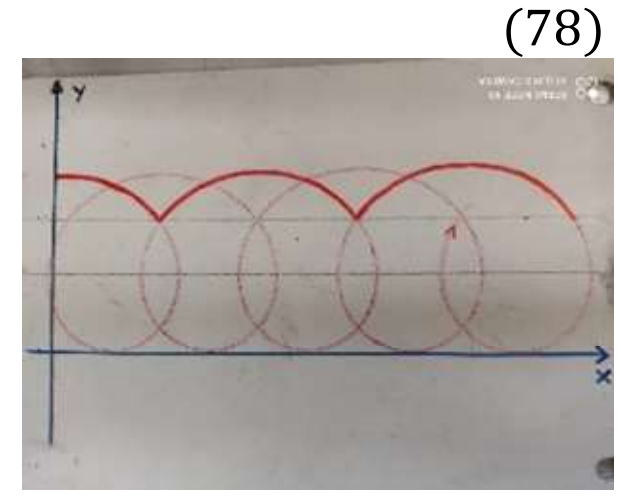

Now we can plot trochoids on the bases of operational function of VSS by using (77) and (78). Therefore, for $\omega = 1$ , $R = 10$ in order to plot a cycloid, it is required $r = 10$ ($r = 5$ for a trochoid on backward sliding mode and $r = 20$ for a trochoid on the forward sliding mode) Fig 6.

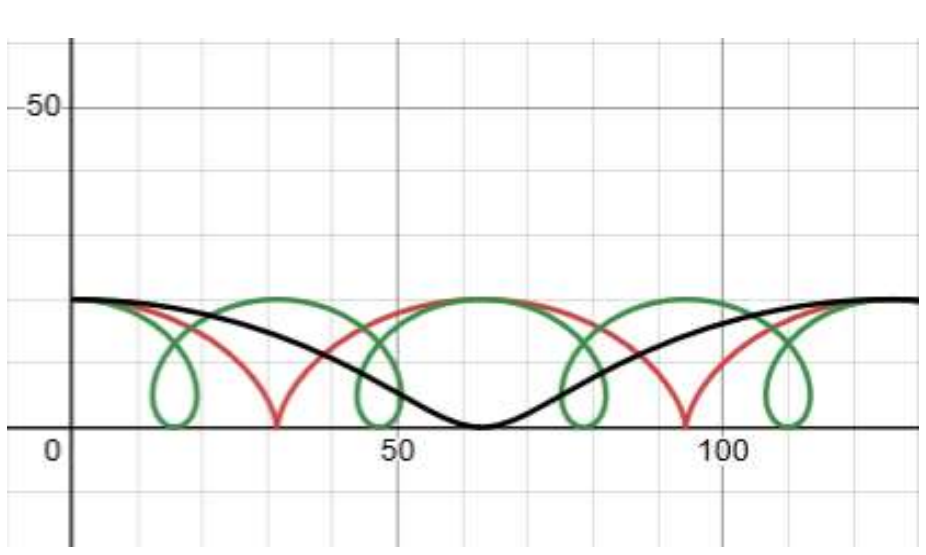

$x(t) = 5t + 10\sin t\ , y(t) = 10 + 10\cos t$

$x(t) = 10t + 10\sin t\ , y(t) = 10 + 10\cos t$

$x(t) = 20t + 10\sin t\ , y(t) = 10 + 10\cos t$

**Fig 6.** Red: Cycloid (pure rolling mode) Green: (backward sliding mode) Black: (forward sliding mode)

Assume that L is the displacement due to pure rolling of gear (along the linear gear) and n is the number of its rotations about symmetry axis. If L' to be the displacement of the tablet (pen holder) due to its pure rolling, then, it follows that

$$L = 2\pi rn \quad (79)$$

$$L' = 2\pi Rn \quad (80)$$

Therefore, $\Delta L = L' - L$ denotes the sliding displacement of the tablet along X axis. Thus

$$\frac{\Delta L}{L} = \frac{R - r}{r} \quad (81)$$

Denotes relative sliding displacement. Thus, $\frac{\Delta L}{L} < 0$ denotes forward sliding and $\frac{\Delta L}{L} > 0$ denotes backward sliding mode. It is obvious from (81) that if $r = R$ there is only pure rolling motion.

## 5.Discussion and Conclusion

Creating trochoid and ellipse through VRCT allows us to introduce a different novel theory for definition and creating these geometrical objects on the bases of meaningful physical concepts. Application of basic physical concepts through VRCT to create geometrical objects refers to establish a strong link between mathematical visions and physical *realities in the world. New mathematical perspective that has been presented in this article proves that all the geometrical objects in the trochoid family and co- centered ellipses can be swept by a definite point on the circumference of virtual circles that are rolling and sliding along other circles!* An outstanding feature of *this article is present a method for deduction and application of theorem1 which is based on a novel idea to simulate pure rolling motion of a circle along another circle. Based on this new vision, it is no longer necessary to consider trochoids as the geometrical objects that are swept by attached points (a meaningless operation) to the pure rolling circles.* Designing and construction MS machine that was used in the course of this research not only provides the possibility of conducting experiments on various physical modeling, but it also provides a set of visible and conceptual experiments that were considered to be impossible. Now these experiments are ready to use for teaching deeply mathematics and physics at universities and colleges. This paper opens a novel, simple and conceptive mathematical perspective based on physical concepts for trochoids and ellipses.